\documentclass[%
reprint,prd,
nofootinbib,
amsmath,amssymb,
aps,superscriptaddress,
]{revtex4-1}

\usepackage{graphicx}
\usepackage{dcolumn}
\usepackage{bm}
\usepackage{xcolor}
\usepackage{mathrsfs}
\usepackage{hyperref}
\hypersetup{
	pdftitle={},
	colorlinks=true,     
	linkcolor=blue,      
	citecolor=teal,      
	filecolor=black,      
	urlcolor=black        
}
\usepackage{cleveref}
\usepackage{amsmath,amssymb}
\usepackage{amsfonts}

\newcommand{\xx}{\mathbf{x}}
\newcommand{\pp}{\mathbf{p}}
\newcommand{\rr}{\mathbb{R}}
\newcommand{\zz}{\mathbb{Z}}
\newcommand{\nn}{\mathbb{N}}
\newcommand{\dd}{\mathrm{d}}
\newcommand{\topo}{{\mathrm{topo}}}
\newcommand{\ener}{{\mathrm{energy}}}
\newcommand{\cl}{{\mathrm{cl}}}

\begin{document}
	
	\preprint{APS/123-QED}
	
	\title{Detecting defect dynamics in relativistic field theories far from equilibrium\\ using topological data analysis}
	
	\author{Viktoria Noel}
 \email{v.noel@thphys.uni-heidelberg.de}
\affiliation{Institute for Theoretical Physics, Heidelberg University, Philosophenweg 16, 69120 Heidelberg, Germany}
 
\author{Daniel Spitz}
\affiliation{Institute for Theoretical Physics, Heidelberg University, Philosophenweg 16, 69120 Heidelberg, Germany}
\affiliation{Max Planck Institute for Mathematics in the Sciences, \mbox{Inselstraße 22, 04103 Leipzig}, Germany}

	\date{\today}
	
	\begin{abstract}

We study nonequilibrium dynamics of relativistic $N$-component scalar field theories in Minkowski space-time in a classical-statistical regime, where typical occupation numbers of modes are much larger than unity. 
In this strongly correlated system far from equilibrium, the role of different phenomena such as nonlinear wave propagation and defect dynamics remains to be clarified. 
We employ persistent homology to infer topological features of the nonequilibrium many-body system for different numbers of field components $N$ via a hierarchy of cubical complexes. 
Specifically, we show that the persistent homology of local energy density fluctuations can give rise to signatures of self-similar scaling associated with topological defects, distinct from the scaling behaviour of nonlinear wave modes.
This contributes to the systematic understanding of the role of topological defects for far-from-equilibrium time evolutions of nonlinear many-body systems.

	\end{abstract}

	
	\maketitle
	
	
	\section{Introduction}\label{sec:1}

Far-from-equilibrium universality can provide a powerful tool to understand the otherwise complex relaxation dynamics of isolated quantum many-body systems.
Specifically, nonthermal fixed points form nonequilibrium attractor solutions to the quantum-dynamical equations of motion, which are characterised by emergent universal self-similar scaling in time~\cite{Berges:2008sr, Walz:2017ffj, Chantesana:2018qsb, Mikheev:2018adp, Schmied:2018upn, Mazeliauskas:2018yef, Li:2022tdz, Mikheev:2022fdl, Heinen:2022ham, Preis:2022uqs, Heller:2023mah}.
This is similar to the universal behaviour of systems close to equilibrium and near criticality~\cite{Hohenberg:1977ym} but without the necessity to fine-tune parameters.
Out of equilibrium, such behaviour has occurred in numerical simulations of both relativistic~\cite{Berges:2008wm, Gasenzer:2011by, Berges:2012us, Berges:2013eia, Ewerz:2014tua, Maraga2015AgingCoarsening, Moore:2015adu, Berges:2016nru, Boguslavski:2019fsb, Mace:2019cqo, Shen:2019jhl} and nonrelativistic~\cite{Nowak:2011sk, Schole:2012kt, Karl:2013kua, Hofmann2014CoarseningDynamicsBinary, Karl:2016wko, Schachner:2016frd, Williamson2016CoarseningDynamics, Deng:2018xsk, Schmied:2018osf, Bhattacharyya:2019ffv, Dolgirev:2019rua, Schmied:2019abm, Gresista:2021qqa, Rodriguez-Nieva:2021okd, Heinen:2022rew, Siovitz:2023ius} models.
Nonthermal fixed points can play a role for dynamics occurring on widely different energy scales, ranging from the early universe~\cite{Micha:2002ey, Micha:2004bv} to collisions of heavy nuclei~\cite{Schlichting:2019abc,Berges:2020fwq} and experiments with ultracold atoms, where they have been observed in recent years~\cite{Prufer:2018hto, Erne:2018gmz, eigen2018universal, Prufer:2019kak, Glidden:2020qmu, Lannig:2023fzf, Huh:2023xso}.
The corresponding universality classes can be surprisingly large, for instance relativistic and nonrelativistic scalar fields forming one class~\cite{PineiroOrioli:2015cpb}, and gauge and scalar fields forming another~\cite{Berges:2014bba, Berges:2019oun}.
Yet, a thorough classification of systems according to their far-from-equilibrium universal behaviour is lacking to date.

Relativistic scalar fields with global $\mathrm{O}(N)$ symmetry provide a paradigm for understanding far-from-equilibrium dynamics, which we focus on in this study.
In particular, such a model allows for the analytic derivation of scaling exponents related to nonthermal fixed points up to anomalous dimensions based on the two-particle-irreducible effective action in large-$N$ expansions~\cite{Berges:2008wm, Berges:2008sr, Scheppach:2009wu, Berges:2010ez, PineiroOrioli:2015cpb, Berges:2016nru, Walz:2017ffj, Chantesana:2018qsb, Mikheev:2018adp}.
Moreover, at early times, for weak couplings and large occupations, the quantum dynamics of the model can be accurately described by means of classical-statistical simulations~\cite{PineiroOrioli:2015cpb, Berges:2015kfa}, which we employ.

Recent experimental and theoretical studies of diverse systems indicate that for a small number of field components different initial conditions can lead to distinct nonthermal fixed points.
They can be related to separate physical mechanisms: self-similar dynamics consistent with descriptions in the limit of many field components~\cite{Prufer:2018hto, Prufer:2019kak}, and topological defects undergoing coarsening dynamics for few field components~\cite{Nowak:2010tm, Gasenzer:2011by, Nowak:2011sk, Schole:2012kt, Karl:2013kua, Ewerz:2014tua, Karl:2016wko, Deng:2018xsk, Schmied:2018osf, Schmied:2019abm, doi:10.1126/science.aat5793, Siovitz:2023ius, Lannig:2023fzf, Huh:2023xso}.
Also for the $\mathrm{O}(N)$ vector model in three spatial dimensions, a number of different, $N$-dependent topological defects are expected to contribute to the dynamics in the infrared~\cite{Moore:2015adu}.
Remarkably, for this model and typical over-occupied initial conditions, many types of equal-time correlation functions merely reveal scaling dynamics consistent with the large-$N$ descriptions, even for low~$N$, and in agreement with nonrelativistic complex $\mathrm{U}(N)$ vector models~\cite{Scheppach:2009wu, PineiroOrioli:2015cpb, Moore:2015adu, Walz:2017ffj, Chantesana:2018qsb, Mikheev:2018adp, PineiroOrioli:2018hst, Boguslavski:2019ecc}.
This includes the momentum-resolved distribution function, which is computed from equal-time two-point correlation functions and forms the basis for many studies of far-from-equilibrium universality.
The investigation of unequal-time two-point correlation functions has revealed differences among the excitation spectra for different $N$~\cite{Boguslavski:2019ecc}.
However, this does not contain clear links to the scaling behaviour related to the coarsening dynamics of topological defects.
A thorough understanding of their role for nonthermal fixed point dynamics can be a prerequisite for the comprehensive classification of universal behaviour far from equilibrium, based on the experimentally and numerically found scaling dynamics.
Hence, there is a need for topologically sensitive and computationally accessible observables.

Topological data analysis (TDA)~\cite{otter2017roadmap, chazal2021introduction} can provide complementary information to correlation functions.
Persistent homology, which is the primary tool from the TDA toolbox, gives access to the topology of a nested hierarchy of complexes constructed from the field data.
It has been successfully applied to describe nonthermal fixed points for two-dimensional nonrelativistic scalar and three-dimensional gluon fields~\cite{Spitz:2020wej, Spitz:2023wmn}, accompanied by a corresponding mathematical analysis~\cite{spitz2020self}.
In the related context of phase transitions, persistent homology allowed for insights into critical phenomena and phase structures~\cite{donato2016persistent, Hirakida:2018bkf, speidel2018topological, santos2019topological, olsthoorn2020finding, Sale:2021xsq, tran2021topological, cole2021quantitative, Kashiwa:2021ctc, Sehayek:2022lxf,  Sale:2022qfn, Spitz:2022tul}.

In this work, we consider the persistent homology of local energy density fluctuations in $\mathrm{O}(N)$ vector models in the classical-statistical regime, starting from over-occupied initial conditions without imprinting defects.
Local energy densities can show clear signatures of dynamically built-up defects as we demonstrate.
In particular, we show that the related Betti number distributions give rise to $N$-dependent topological structures reminiscent of the classification of defects in condensates of relativistic $\mathrm{O}(N)$ scalar fields.
The time dependence of the related length scales indicates topological dynamics consistent with phase-ordering kinetics~\cite{Bray:1994zz}.

The remainder of this paper is structured as follows: \Cref{sec:2} describes the lattice simulations, along with a description of topological defects in $\mathrm{O}(N)$ theories and their presence in local energy densities. 
\Cref{sec:energy} introduces persistent homology and discusses the signals of topological defects in Betti number distributions.
Moreover, the latter are interpreted in light of the coarsening dynamics of topological defects and energy transport.
Finally, \Cref{sec:discussion} provides a conclusion. 
 
\section{$\mathrm{O}(N)$ vector model and topological defects}\label{sec:2}
We consider a relativistic $\mathrm{O}(N)$-symmetric scalar field theory with field variables $\phi_a(t,\xx)$, $a=1,...,N$, in $d=3$ spatial dimensions with classical action
\begin{equation}\label{eq:action}
		S[\phi]=\int_{t, \xx}\left[\frac{1}{2} \partial^\mu \phi_a \partial_\mu \phi_a-\frac{m^2}{2} \phi_a \phi_a-\frac{\lambda}{4 ! N}\left(\phi_a \phi_a\right)^2\right],
\end{equation}
where $\int_{t, \xx} \equiv \int \mathrm{d} t \int \mathrm{d}^3 \xx$, summation over repeated indices is implied, $m$ is the bare mass and $\lambda$ is the coupling constant. 
The action \eqref{eq:action} is invariant under global $\mathrm{O}(N)$ rotations acting on the internal field components indexed by $a$: $\phi_a\mapsto R_{ab} \phi_b$ for $R\in \mathrm{O}(N)$.

\subsection{Lattice simulations}\label{sec:lattice}
For highly occupied systems at not too late times and weak couplings, the quantum dynamics can be accurately mapped to a classical-statistical field theory (truncated Wigner approximation)~\cite{Son:1996zs, Aarts:2001yn, Smit:2002yg, Berges:2007ym, Polkovnikov:2009ys, Berges:2013lsa}.
In classical-statistical simulations, one samples over initial conditions and each realisation is evolved according to the classical equation of motion. 
Expectation values of observables are obtained by averaging over their evaluations for the classical trajectories. 
Real-time classical-statistical simulations have been extensively used to study the dynamics of scalar fields within closed quantum systems in corresponding regimes of applicability~\cite{Aarts:2001yn, Mueller:2002gd, Skullerud:2003ki, Jeon:2004dh, Arrizabalaga:2004iw, Berges:2013lsa, PineiroOrioli:2015cpb, Boguslavski:2019ecc}.

We consider over-occupied box initial conditions in momentum space for the scalar fields.
More specifically, with the statistical two-point correlation function, defined as
\begin{equation}
    F(t,t',\xx-\xx') = \frac{1}{2N}\langle \phi_a(t,\xx)\phi_a(t',\xx') + \phi_a(t',\xx')\phi_a(t,\xx)\rangle
\end{equation}
for spatially homogeneous scenarios, occupation numbers can be defined as
\begin{equation}
    f(t,\pp) = \sqrt{F(t,t',\pp)\partial_t\partial_t' F(t,t',\pp)}\big|_{t=t'},
\end{equation}
where $F(t,t',\pp) = \int_{\Delta\xx} F(t,t',\Delta \xx) \exp(-i\pp\Delta\xx)$ and $\int_{\Delta\xx}\equiv \int \dd^3 \Delta \xx$.
Then field configurations $\phi_a(t=0,\xx)$ are sampled with large Gaussian fluctuations up to a characteristic momentum scale $Q$, described by 
\begin{equation}\label{EqOccTimeZero}
			f(t=0, \mathbf{p})=\frac{Nn_0}{\lambda}\Theta(Q-\mathbf{p}),
\end{equation}
with zero macroscopic fields $|\phi(t=0,\xx)| = 0$.
The initial field configurations are time-evolved according to the equation of motion following from the action \eqref{eq:action}.
For more details we refer to~\cite{PineiroOrioli:2015cpb, Berges:2015kfa}.

We consider lattice simulations for $N_s^3=512^3$ spatial lattice sites and use a leapfrog solver with $Q a_s =0.8$, where $a_s$ is the spatial lattice constant, temporal lattice spacing $\dd t = 0.1 \, a_s$, coupling $\lambda=0.1$, initial amplitude $n_0=125$ and renormalised mass squared $M^2 = 4 Q^2$, which is used in an iterative procedure to fix the bare mass $m$ as described in~\cite{Berges:2015kfa}.
The comparably large mass suppresses fluctuations on smaller length scales, which will allow us to reveal the presence of topological features associated with defects.
In persistent homology as introduced below, the topological features appear less pronounced for smaller values of the mass, see \Cref{AppendixMassDiscussion}.
For later use we denote the spatial lattice by $\Lambda_s:=\{0,a_s,\dots,(N_s-1)a_s\}^3$.

In this work we focus on the structures visible in local energy density fluctuations around the mean energy density.
The local energy density corresponding to the action \eqref{eq:action} is:
\begin{equation}\label{eq:T00}
			T^{00}(t,\xx)= \frac{1}{2} \pi^2_a + \frac{1}{2} (\nabla \phi_a)^2 + \frac{m^2}{2} \phi_a \phi_a + \frac{\lambda}{4!N}(\phi_a \phi_a)^2,
\end{equation}
where $\pi_a = \partial_t \phi_a$ and space-time arguments have been suppressed on the right-hand side.
Central spatial derivatives are employed on the lattice.

We consider a single classical-statistical realisation in this work, based on the self-averaging property often encountered for observables in classical-statistical simulations.
The latter also holds approximately for the later introduced Betti numbers due to the large number of contributing features.%
\footnote{For mathematical details on self-averaging related to persistent homology (ergodicity in persistence) we refer to~\cite{spitz2020self}.}
In fact, the Betti numbers are expected to scale proportionally to the system volume for sufficiently large lattices, based on mathematical theorems~\cite{hiraoka2018limit, spitz2020self}.
Specifically, we have verified insensitivity of Betti numbers after division by the lattice volume for $N_s=128, 256, 512$.
We have also numerically verified their insensitivity to variations of the lattice spacing upon comparison with simulations for $Qa_s=0.6$ and ${Qa_s= 1.2}$.

\subsection{Topological defects in relativistic $\mathrm{O}(N)$ theories}\label{sec:defects}

\subsubsection{Topological defects in condensates}\label{sec:defect}
The chosen initial condition sets high occupation numbers up to the momentum scale $Q$. 
As discussed in previous works~\cite{Berges:2012us, PineiroOrioli:2015cpb}, the particle number redistributes towards the infrared via an inverse particle number cascade, which is a consequence of transient approximate particle number conservation.
In this dynamical process, a condensate forms in the zero mode, which is initially absent and results from increasing occupancies in the deep infrared.
Simultaneously, long-range order gradually builds up~\cite{Moore:2015adu}.%
\footnote{By causality, the formation of a condensate at finite evolution times requires a finite system volume.
For infinite system volumes, modes in the deep infrared can mimic the dynamics of a condensate, but lack the related system-scale long-range order~\cite{Moore:2015adu}.}

Considering the ordering dynamics of the condensate, the phase space of spatial zero modes of the (Fourier-transformed) field variables $\tilde{\phi}_a(t,\pp) = \int_\xx \phi_a(t,\xx)\exp (-i\pp\xx)$, formed by $(\tilde{\phi}_a,\partial_t\tilde{\phi}_a)_{a=1,\dots,N}$ with $\tilde{\phi}_a\equiv \tilde{\phi}_a(t,\textbf{p}=\textbf{0})$, is of relevance.
Approximate particle number conservation and energy minimisation provide two constraints for the realised condensate configurations~\cite{Moore:2015adu}.
Taking these into account, we denote the physically accessible condensate phase space by $\mathscr{C}_N$, which depends nontrivially on the number of field \mbox{components $N$}.
In particular, the topology of $\mathscr{C}_N$ can be nontrivial, so that topologically nontrivial configurations (defects) can occur.
These have been classified in~\cite{Moore:2015adu}, which we review in detail in \Cref{AppendixA}.
Specifically, the condensate can feature string defects (vortex lines) for $N=1,2,3$, domain walls for $N=2$ and monopoles for $N=4$.
For three spatial dimensions no other defects are expected, and condensates are defect-free for $N\geq 5$.

Note that defects can be dynamically generated during the evolution of the (classical) equation of motion, and are not explicitly part of the initial conditions under consideration.
As described later in this work, defects typically annihilate each other with time via related coarsening dynamics~\cite{Bray:1994zz}, such that their number decreases.
On longer time scales than considered here, the condensate itself is expected to decay again due to number-changing processes in the relativistic theory~\cite{Berges:2012us}, suppressing topological defects as well.

\begin{figure*}[!htbp]
			
	\includegraphics[scale=0.6]{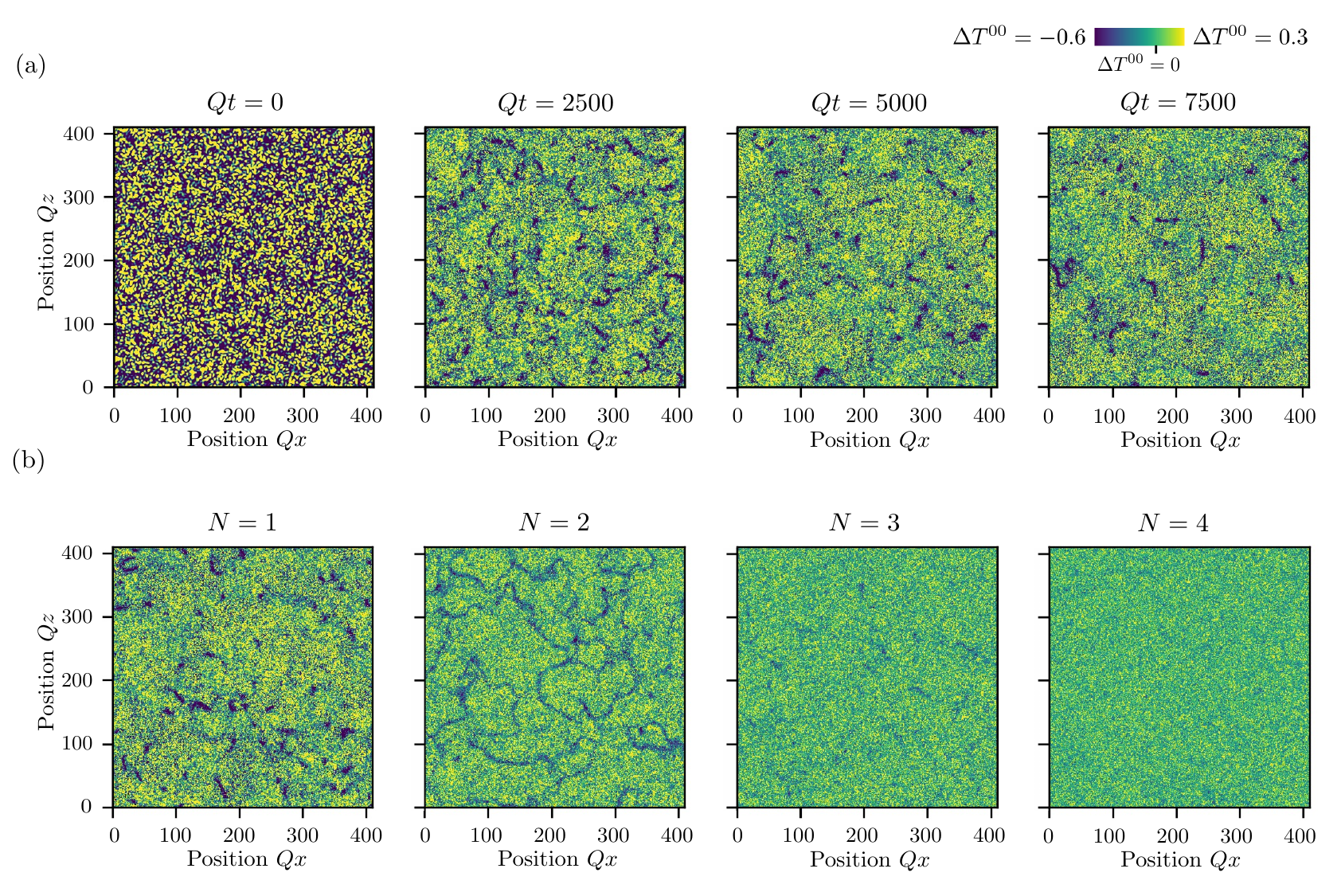}%
			
	\caption{\label{fig:energies1}%
	Two-dimensional slices of local energy density fluctuations around their mean value, given by $\Delta T^{00} = (T^{00}-\bar{T}^{00})/\bar{T}^{00}$. 
    (a) Snapshots at $N=1$ for different times, and (b) snapshots for different $N$ at time $Qt=5000$, showing signatures of string defects for $N=1,2,3$ and of domain walls for $N=2$. }
\end{figure*}

\subsubsection{Observation of defects in energy density fluctuations}\label{defectindens}
We probe topological defects via their signatures in local energy density fluctuations around their mean values, given by
\begin{equation}
    \Delta T^{00}(t,\xx) := \frac{T^{00}(t,\xx) - \bar{T}^{00}}{\bar{T}^{00}},
\end{equation}
where $\bar{T}^{00} := (1/N_s^3)\sum_{\xx\in \Lambda_s} T^{00}(t,\xx)$ is time-independent due to energy conservation.
In \Cref{fig:energies1}(a) we display two-dimensional slices of the time-evolving energy density fluctuations for $N=1$. 
As time elapses, the initial fluctuations in local energy densities become gradually more homogeneous, with string-like structures distinctively emerging at early times and minimal energy densities.
We associate these with string defects, whose number appears to decline with time.

Comparing the two-dimensional snapshots of energy density fluctuations for $N=1,2,3,4$ at a fixed time ($Qt=5000$), displayed in \Cref{fig:energies1}(b), we observe similar string-like structures for $N=2$ and $N=3$.
Moreover, for $N=2$ we can also see indications of emerging domains separated by domain walls.%
\footnote{These have been observed before in other observables, see e.g.~\cite{Gasenzer:2011by, Berges:2017ldx}.}
These appear at locally low energy densities as closed loops in the snapshots and can be discriminated from the open elongated minima attributed to string-like structures. 
For $N=4$ and \mbox{higher $N$} (not shown), the local energy densities become gradually more homogeneous without any distinct string- or domain-like structures forming, even when running the simulations for comparably long times up to $Qt=50000$.
Such defect structures agree with the classification outlined in \Cref{sec:defect}, albeit with monopoles not seen for $N=4$.
Monopoles form point-like local minima in energy densities, similar to the many other fluctuations present in the system.
The lattice configurations can therefore still contain monopoles, but might be indistinguishable from configurations without monopoles based on our methods.

The emergence of defect structures at minima in local energy density fluctuations can be heuristically understood based on energetic considerations.
Defects locally minimise potential energy densities due to necessary zero-crossings in field amplitudes between oppositely-signed domains, at a string or within a monopole core, which manifest as zero local potential energy densities.
Kinetic energy densities require more careful considerations.
Defects move slowly in comparison to typical time scales associated with the inverse particle cascade, resulting in small contributions by $\sum_a \pi_a^2$ to $T^{00}$ in \Cref{eq:T00}.
Moreover, spatial gradients in directions tangent to the curves formed by local energy density minima due to string defects or tangent to the corresponding surfaces for domain walls are naturally small, as in these directions local energy densities do not change much (see \Cref{fig:energies1}).
In normal directions defects come with related characteristic (healing) length scales.
These can be comparably large, leading to a suppression of normal gradients as well, such that kinetic energy densities of defects are locally suppressed along with potential energy densities.
We have observed this in our simulations for both kinetic and potential energy densities (not shown).

\section{Persistent homology of energy density fluctuations}\label{sec:energy}

Persistent homology provides a method to calculate scale-dependent topological structures from data along with measures of their persistence.
In \Cref{sec:cube} we present the concept of persistent homology for the investigated local energy density fluctuations. 
In \Cref{sec:phdef} the persistent homology results are discussed in light of the previous discussion of topological defects in condensates.
\Cref{sec:scaling} considers the time dependence of Betti number distributions in relation to coarsening dynamics, while \Cref{sec:energytransport} provides an examination of its connection to energy transport towards the ultraviolet.

\subsection{Persistent homology of energy density fluctuation sublevel sets}\label{sec:cube}

\subsubsection{Cubical complexes}
		
In this work, persistent homology is computed for cubical complexes~\cite{wagner2012efficient}, which is ideally suited for data in pixel format.
Here, we focus on a short, intuitive introduction, while \Cref{AppendixMathDetailsComplexesPersHom} provides mathematically more concise constructions.
For more elaborate mathematical introductions to TDA we refer to~\cite{otter2017roadmap, chazal2021introduction}.

\begin{figure*}
	\includegraphics[width=0.8\textwidth]{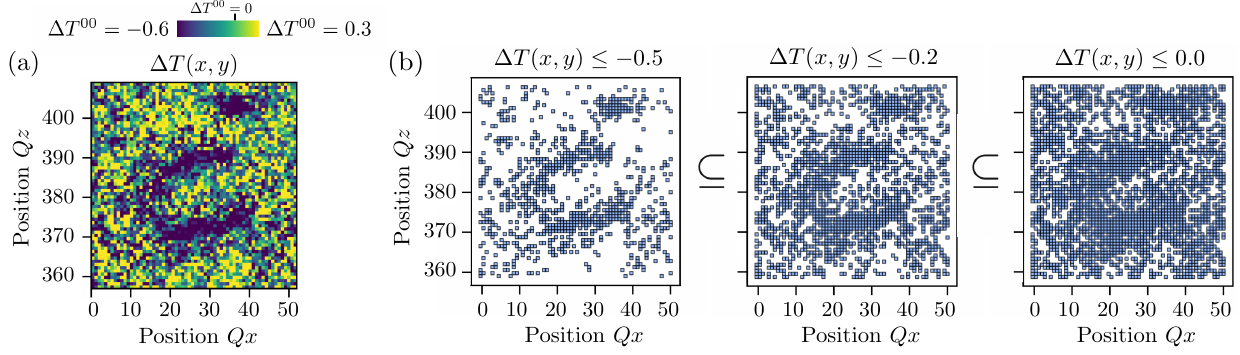}
	\caption{\label{fig:sublevel}%
	Pixelisation of lattice sublevel sets leads to a filtration of cubical complexes. 
    (a): $64 \times 64$ pixels excerpt from the top-left corner of the $N=1$ energy density fluctuation slice shown in \Cref{fig:energies1}(b).
    (b): Three nested cubical subcomplexes corresponding to the indicated sublevel sets of the image of (a), for filtration parameters $\nu\equiv \Delta T^{00}\leq -0.5, -0.2, 0.0$ from left to right.}
\end{figure*}

A cubical complex is a collection of cubes of different dimensions, closed under taking boundaries. 
For instance, the boundary of a 3-cube is the union of all its six faces, i.e., boundary squares, the boundary of a 2-cube (square) is the union of its four boundary edges, the boundary of a 1-cube (edge) consists of its two endpoints, and the boundary of a 0-cube (point) is empty.
The full cubical complex for the spatial lattice under consideration consists of a 3-cube for each lattice site, with a lattice site located at each cube's center.
For each 3-cube all boundary cubes of lower dimensions are included in the full cubical complex.

Subcomplexes of the full cubical complex can be used to describe sublevel sets of functions on the spatial lattice. 
Intuitively, for a given function these are constructed by including a 3-cube in the subcomplex whenever the corresponding function value is below a chosen filtration parameter.
While this fixes the filtration parameters when 3-cubes enter subcomplexes, the filtration parameters for lower-dimensional cubes are set inductively (the lower star filtration, see \Cref{AppendixMathDetailsComplexesPersHom}).
This way the subcomplex becomes indeed a cubical complex for every filtration parameter, which can be seen as a pixelisation of the corresponding lattice function sublevel set.
In particular, a nested sequence of cubical complexes is formed upon increasing the filtration parameter.
This procedure is illustrated in \Cref{fig:sublevel} for the example of an excerpt of the two-dimensional slice of the local energy density fluctuations shown in \Cref{fig:energies1}(b) ($N=1$).
While the two-dimensional slice is for illustrative purposes only, the analysis is carried out in all three spatial dimensions.
For a given function on the lattice such as the image excerpt given in \Cref{fig:sublevel}(a), the cubical complexes corresponding to the sequence of sublevel sets can resemble the structure of minima, see \Cref{fig:sublevel}(b). 
In particular, note the persistence of the horseshoe-like accumulation of squares across filtration parameters.

More formally, the evaluation of local energy density fluctuations for a given classical-statistical realisation $\phi(t,\xx)$ at time $t$ provides a real-valued function $\Delta T^{00}(t,\cdot)$ on the spatial lattice $\Lambda_s$.
Its sublevel sets are defined as 
\begin{align}
	M_{\Delta T^{00}}(t,\nu):=&\;(\Delta T^{00}(t,\cdot))^{-1}(-\infty, \nu]\nonumber\\
 =&\;\{\xx \in\Lambda_s\,|\,  \Delta T^{00}(t,\xx) \leq \nu\}.
\end{align}
The described pixelisation procedure leads to the cubical complexes $\mathcal{C}_{\Delta T^{00}}(t,\nu)$, which are the cubical complexes of interest in this work.
They form a filtration of the full cubical complex, i.e., a nested sequence of subcomplexes of the full cubical complex with
\begin{equation}
    \mathcal{C}_{\Delta T^{00}}(t,\nu) \subseteq \mathcal{C}_{\Delta T^{00}}(t,\mu)
\end{equation}
for all $\nu\leq \mu$.
For $\nu$ smaller than the minimum value of $\Delta T^{00}(t,\cdot)$ the cubical complex $\mathcal{C}_{\Delta T^{00}}(t,\nu) $ is empty, for $\nu$ larger or equal the corresponding maximum value the full cubical complex is recovered.

		\begin{figure}
			\includegraphics[scale=0.49]{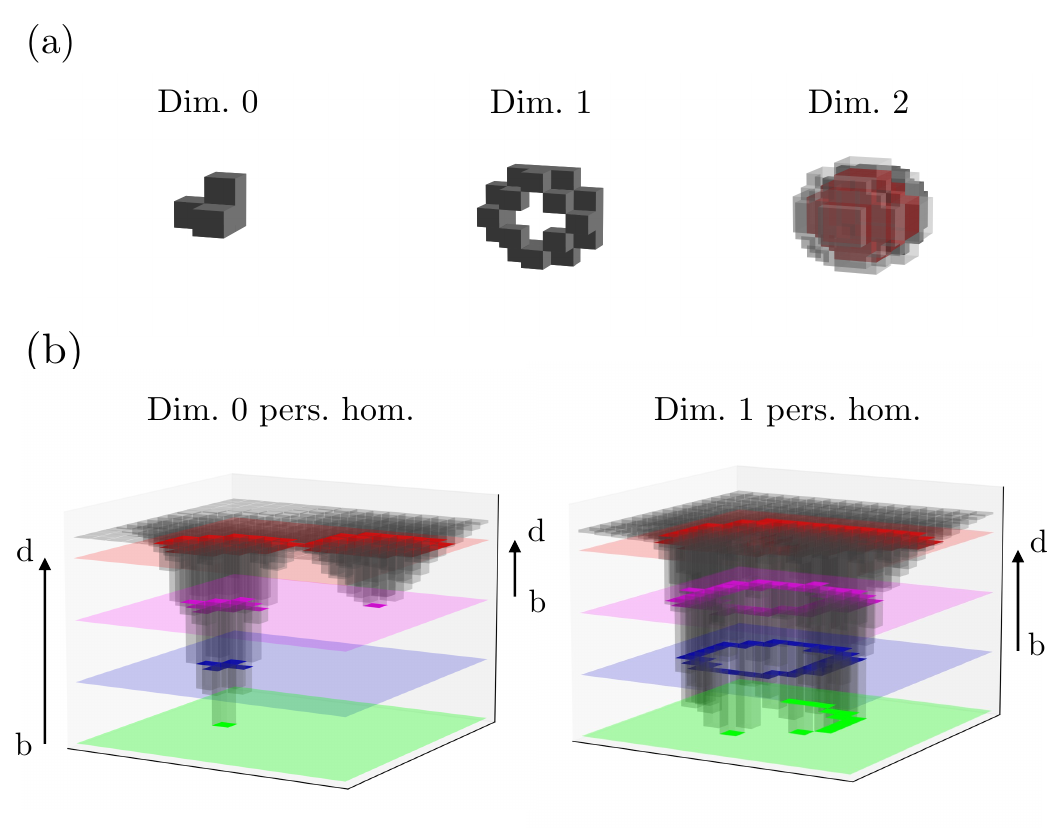}
			\caption{\label{fig:cubes}(a): Dimension-0, -1 and -2 homology classes from left to right, built with cubical complexes. Dimension-2 homology classes are enclosed volumes, indicated in red. (b): Dimension-0 and -1 holes in cubical complexes for different sublevel sets, giving rise to persistent homology classes.}
		\end{figure}

\subsubsection{Persistent homology: holes in complexes}\label{sec:holes}

The full cubical complex of the three-dimensional lattice contains a 3-cube for each spatial lattice point. 
However, as energy density fluctuation values are swept through from the lowest to the highest, in general, the cubical complexes $\mathcal{C}_{\Delta T^{00}}(t,\nu)$ do not contain a cube for every spatial lattice point.
Holes of different dimensions can appear, which are described by homology groups. 
For cubical complexes, such holes are illustrated in \Cref{fig:cubes}(a).
Connected components are described by dimension-0 homology classes, dimension-1 homology classes describe planar-like holes (which in three dimensions can also be viewed as tunnels), and dimension-2 homology classes describe enclosed volumes. 

Within the filtration, holes are born at a birth parameter $b$ and die again with death  parameter $d$, possibly deforming as the (energy density fluctuation) filtration is swept through, i.e., being present in the filtration for the filtration parameter interval $[b,d)$. 
The persistence $p=d-b$ is a measure of the dominance of a feature in the filtration. 
This is illustrated in \Cref{fig:cubes}(b), where two landscapes of example functions with distinct minima are displayed.
The left function contains two distinct dips. 
As the sublevel set filtration is swept through, the cubical complex is empty as long as the filtration \mbox{parameter $\nu$} is less than the minimum value of the function. 
As $\nu$ is increased, a dimension-0 homology class is born with birth parameter $b=\nu$ when the minimum value of the function is reached (green plane). 
Further increasing $\nu$ (blue plane), the single 2-cube turns into a set of 2-cubes, leaving this homology class unchanged. 
When the value of $\nu$ reaches the pink plane, a second dimension-0 homology class is born (on the right) corresponding to the second dip in the function. 
The two dimension-0 homology classes merge into one at the red plane, at which point the first homology class dies with death parameter $d=\nu$, and only the second one survives.

Turning to the right function in \Cref{fig:cubes}(b), a few dimension-0 homology classes are born (green plane), merging to form a dimension-1 homology class when $\nu$ is increased to the level of the blue plane, represented by the circular structure in the complex surrounding a hole.
Further increasing $\nu$, the homology is unchanged at the pink plane and the dimension-1 hole only disappears somewhere between the pink and red planes when it is fully filled by 2-cubes. 

From the cubical complexes $\mathcal{C}_{\Delta T^{00}}(t,\nu)$, the different dimension-$\ell$ homology groups $H_{\ell}(\mathcal{C}_{\Delta T^{00}}(t,\nu))$ can be computed. 
For three spatial dimensions, the homology groups are generally nontrivial for $\ell=0,1,2$, while the dimension-3 homology group only captures the toroidal lattice topology itself; all higher homology groups are trivial.
Their dimensions, called Betti numbers, specify the number of independent dimension-$\ell$ holes (homology classes):
	\begin{equation}
			\beta_{\ell}(t,\nu):=\operatorname{dim} H_{\ell}(\mathcal{C}_{\Delta T^{00}}(t,\nu)).
	\end{equation}
We focus on Betti numbers as an informative persistent homology descriptor in this work.
		
Persistent homology has a number of useful features.
Notably, it is stable against small perturbations of the input~\cite{otter2017roadmap, chazal2021introduction} (local energy density values in this case) and well-defined large-volume asymptotics exist for suitable persistent homology descriptors such as Betti numbers, including notions of ergodicity~\cite{hiraoka2018limit, spitz2020self}. 
We compute the persistent homology of cubical complexes with $\zz_2$ coefficients and periodic boundary conditions using the open source TDA library GUDHI~\cite{maria2014gudhi}.

To summarise, the persistent homology of sublevel sets can be particularly sensitive to the extended structures formed by local minima and saddle points.
We utilise this to investigate the structure of minima in energy density fluctuations, for which nontrivial defect dynamics is expected, based on the discussion in \Cref{defectindens}.

\begin{figure*}
\includegraphics[scale=0.64]{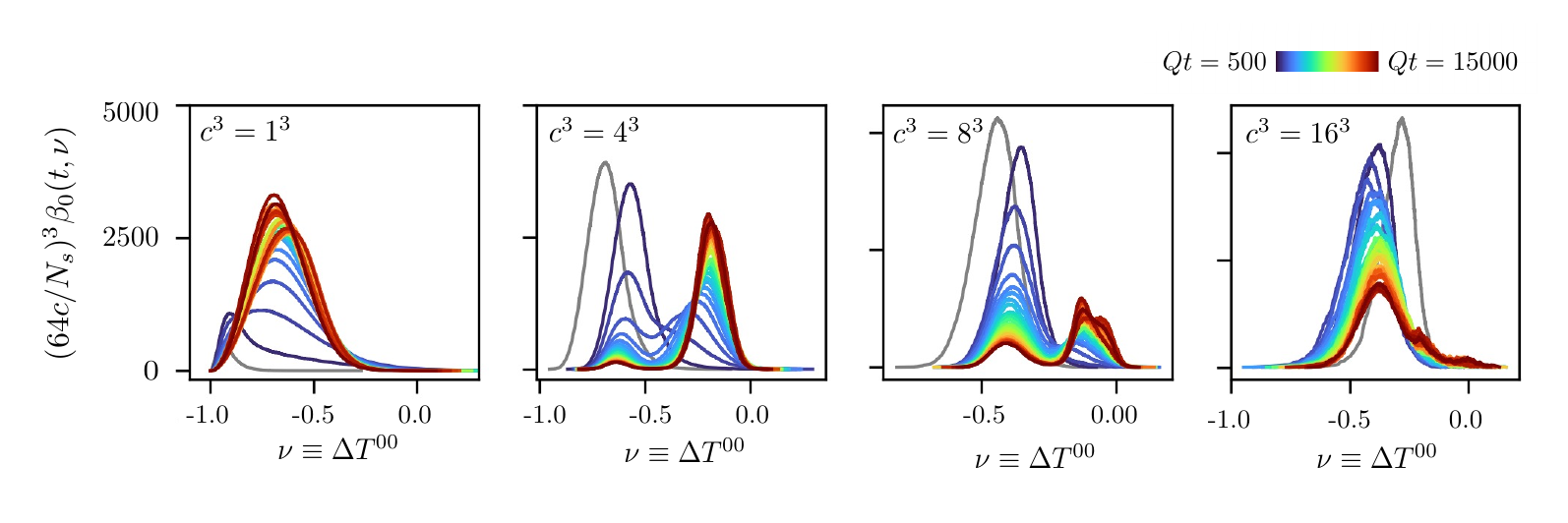}
\caption{\label{fig:coarse}	Dimension-0 Betti numbers for $\Delta T^{00}$ sublevel sets with local energy densities averaged over nearby blocks of every $c^3=1^3,4^3,8^3$ and $16^3$ lattice sites (from left to right). 
Betti numbers have been normalised to $64^3$ lattice size in accordance with their volume-scaling.
Betti numbers at initial time $Qt=0$ are shown in grey.
}
\end{figure*}

\subsubsection{Blockwise averaging lattice functions}

By construction, the persistent homology of cubical complexes does not contain spatial metric information. 
However, this also means that persistent homology computed from functions on the lattice does not differentiate between different length scales and therefore lattice artefacts can enter the analysis. 
By blockwise averaging (coarsening) the local energy densities on the lattice over blocks of $c^3$ points for $c>1$, both these artefacts and, partly, ultraviolet fluctuations are removed.
This can pronounce contributions related to the infrared dynamics including topological defects, depending on the parameter $c$.
We emphasise that the blockwise averaging procedure does not affect the dynamical evolution, since it is only applied \emph{a posteriori} to the lattice configurations as part of the persistent homology read-out.

This is demonstrated in \Cref{fig:coarse} for the case of $N=1$ with dimension-0 Betti numbers. 
The number of persistent homology classes typically decreases as we average blockwise, since the number of lattice sites decreases accordingly.
Betti numbers are expected to scale proportionally to system volume for sufficiently large lattices~\cite{hiraoka2018limit, spitz2020self}, so that we can account for this by volume-normalising to a certain number of lattice sites, here chosen to be $64^3$.%
\footnote{We later focus on blockwise averaging by a factor of $8$, so the effective lattice size for the persistent homology analysis is $(512/8)^3=64^3$.}
Without blockwise averaging, we see that the number of connected components specified by dimension-0 Betti numbers increases at earlier times.
Averaging local energy densities over every $4^3$ lattice blocks, we notice a twofold peak structure emerging: connected components accounting for the left peak decrease in numbers, while the right peak grows with time.
Increasing to averaging over every $8^3$ blocks, the left peak appears more pronounced, while the right peak decreases in height.
This behaviour persists when coarsening by a factor of $16$ in total in each direction.

Concerning the physical interpretation, we notice that the initially large occupations give rise to dynamics towards the infrared and growing wavelengths of correspondingly dominating modes as can be seen in the occupation number distributions, which are comparable to~\cite{PineiroOrioli:2015cpb} (not shown here).
The height of the left peak in dimension-0 Betti numbers decreases over time for energy densities averaged over blocks of $4^3,8^3$ or $16^3$ nearby lattice sites, which implies an increase in the average distance between the related connected components.
This indicates that dimension-0 Betti numbers can probe the infrared dynamics for blockwise averaged configurations.
In particular, the coarsening dynamics of topological defects merging is typically also accompanied by a growth of respective characteristic length scales~\cite{Bray:1994zz}.
In the following, we show results for local energy densities blockwise averaged over cubes of $8^3$ neighbouring lattice points.
When quantitatively discussing the time dependence of Betti numbers below, it is investigated if the observed phenomena are stable with respect to further blockwise averaging.

	\begin{figure*}
		
		\includegraphics[scale=0.6]{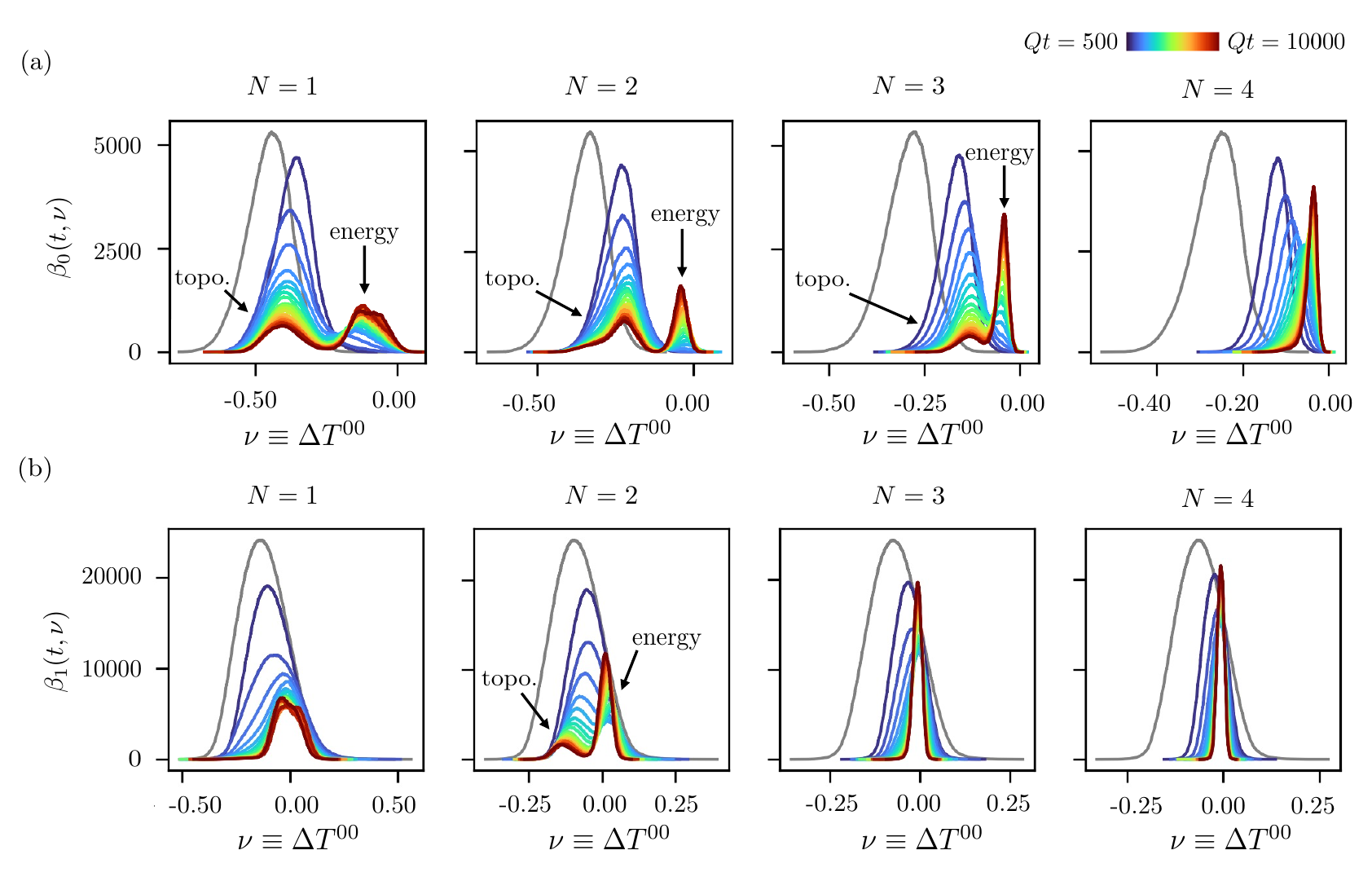}%
		
		\caption{\label{fig:dim0Betti}%
			(a) Dimension-0 and (b) dimension-1 Betti numbers for $\Delta T^{00}$ sublevel sets and $N=1$ to $N=4$ field components (from left to right), with averaging over blocks of $8^3$ lattice sites. 
   Betti number distributions at initial time $Qt=0$ are shown in grey. 
   For Betti number distributions with twofold peak structures we label the peak associated with topological defects by ``topo.~", and the peak corresponding to energy transport by ``energy".
   }%
	\end{figure*}
 
\subsection{Topological defects in Betti numbers}\label{sec:phdef}

As shown in \Cref{fig:dim0Betti}(a), upon investigating the time-evolving dimension-0 Betti numbers of local energy density fluctuations from $N=1$ to $N=4$ with averaging over blocks of $8^3$ lattice sites, we observe two distinct peaks appearing for $N=1,2,3$, while for $N=4$ and higher $N$ (not shown) there is only a single peak. 
Starting from the initial conditions (displayed in grey), for all $N$, the single peak in dimension-0 Betti numbers first moves to larger filtration parameters at early times, i.e., local minima shift to larger energy density values.
For times larger than $Qt=500$, the peak position stays approximately constant for $N= 1,2,3$, while the overall number of dimension-0 structures decreases with time.
For $N=1,2,3$, the single peak splits up into two distinct peaks, the left one decreasing, the right one increasing in height.
For $N=4$, no split-up happens, and the peak first decreases, then increases in height, and continuously shifts to larger filtration parameters.

Since dimension-0 Betti numbers count independent connected components, a decline in peak height implies that average length scales associated with the component configurations increase, which correspond to local energy density minima in the field configurations.
Heuristically, we may thus associate the left peak and its decrease as a signature of dynamics towards larger length scales.
Notably, for $N=1,2,3$, the connected components making up the left peak appear for those $\Delta T^{00}$-values where we have observed defects in the two-dimensional snapshots of $\Delta T^{00}$ in \Cref{fig:energies1}(b).
This indicates that in terms of specific field configurations it is primarily the defects which dominate the declining left peaks in Betti numbers.
Accordingly, the decrease in peak height can signal their coarsening dynamics, which we discuss in more detail later, see \Cref{sec:scaling}.
Vice versa, Betti numbers increasing with time as for the right peak for $N=1,2,3$, and for $N=4$ for later times $Qt\gtrsim 4000$ implies refinement dynamics of the connected components, i.e., an increasing number of local minima appears. 
Corresponding length scales shrink in time.
The peak shifting towards $\Delta T^{00}=0$, this indicates an ongoing homogenisation process of energy densities and can be suggestive of the transport of energy towards towards smaller length scales (larger momenta), as we analyse in more detail in \Cref{sec:energytransport}.

Likewise, the analysis of dimension-1 Betti numbers shows qualitatively similar results for all $N$, but with only a single peak structure that decreases in amplitude, except for $N=2$, for which two peaks appear.
This is displayed in \Cref{fig:dim0Betti}(b). 
More precisely, we notice a height decline of the peak for $N=1,3,4$, which turns into an increase at later times.
For $N=2$, the behaviour is different: the height of the initial peak only decreases, and the single peak splits up into two peaks, where the right one forms at larger energy density values corresponding to $\Delta T^{00}\simeq 0$ and increases in height with time.
The dimension-2 Betti numbers agree qualitatively for all $N$ and do not come with twofold peak structures (not shown). 

Still, a decline in dimension-1 Betti numbers indicates that length scales associated with the configurations of holes increase in time as for the dynamics towards the infrared.
In particular, this applies to the left peak for $N=2$, which constantly decreases in height and appears at energy densities, for which we have inferred the presence of defects from \Cref{fig:energies1}(b).
Similarly to the previous discussion of dimension-0 Betti numbers, at later times, the increase in height for most peaks in dimension-1 Betti numbers together with their shift towards $\Delta T^{00} =0$ can be indicative of refinement dynamics leading to a homogenisation of local energy densities.

To summarise, we observe potentially defect-related peaks in the dimension-0 Betti numbers shown in \Cref{fig:dim0Betti}(a) for $N=1,2,3$ but not for $N=4$, and only for $N=2$ in the dimension-1 Betti numbers displayed in \Cref{fig:dim0Betti}(b).
This is in line with the classification of topological defects for condensates outlined in \Cref{sec:defect} and reviewed in detail in \Cref{AppendixA}, provided that the left peak in dimension-0 Betti numbers is primarily due to strings and the left peak in dimension-1 Betti numbers is mostly due to domain walls.
Indeed, this appears well-motivated from 
the structure of the defects along with fluctuations on top as inferred from the local energy density snapshots displayed in \Cref{fig:energies1}.
For instance, closed strings manifest as loop-like minima in local energy densities and would thus naively appear as distinct dimension-1 persistent homology classes in the sublevel sets for correspondingly low filtration parameters (cf.~also the discussion at the end of \Cref{defectindens}).
Yet, smaller-scale fluctuations on top result in a landscape of local minima and maxima, which adds to this and may interrupt the clean loop-like minima in energy densities.
Accordingly, a closed string would not appear as a distinct dimension-1 homological feature anymore, but as a multitude of dimension-0 structures, which in addition cannot be distinguished from open strings.
Similarly, domain walls would naively show up as empty volumes in energy density sublevel sets for correspondingly small filtration parameters, which would give rise to distinct dimension-2 persistent homology classes.
The addition of smaller-scale fluctuations can yield local energy density maxima, which then pierce the enclosed volumes in the sublevel sets, yielding only scaffold-like networks which surround the empty volumes.
These do not give rise to distinct dimension-2 persistent homology classes anymore, but to an abundance of dimension-1 features. 
The same reasoning applies to more exotic configurations of topological defects such as strings pinned to domain walls~\cite{Karl:2013kua}, which our persistent homology analysis cannot distinguish from the domain walls themselves.

	\begin{figure}
		\includegraphics[scale=0.66]{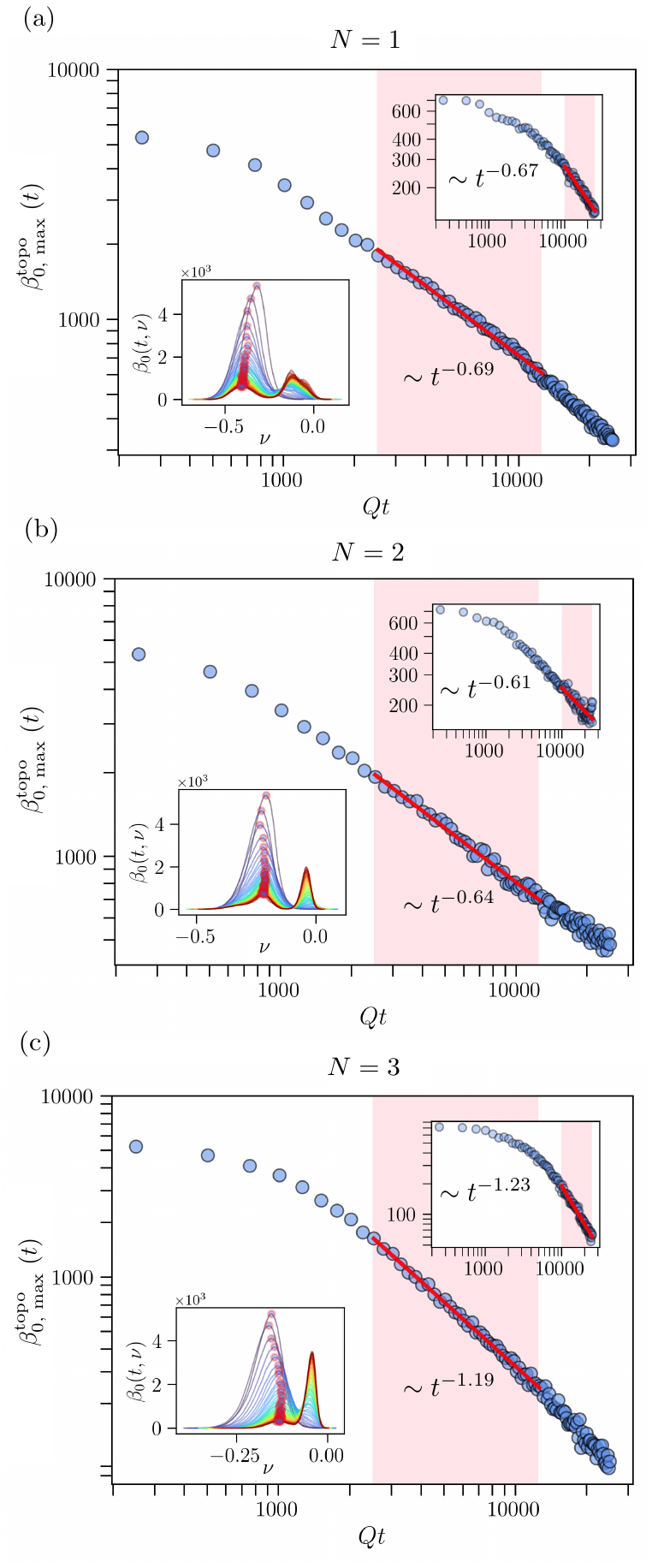}
		\caption{\label{fig:dim0p2} Temporal scaling of the left dimension-0 Betti number distribution peak values for (a) $N=1$, (b) $N=2$ and \mbox{(c) $N=3$}, with averaging over blocks of $8^3$ lattice sites.
  The bottom insets show the Betti number distributions, where the red circles indicate the peak positions. 
  The top insets show peak values for averaging over every $16^3$ blocks.
  The power law fits are based on the data in the red shaded $Qt$ ranges. }
		\label{fig:scaling}
	\end{figure}

\subsection{Signatures of coarsening dynamics}\label{sec:scaling}
On the time scales under consideration, the over-occupied initial conditions lead to nonthermal fixed point dynamics as characterised by dynamical self-similar scaling. 
The temporal dependence of the distribution function is restricted to spatial rescalings by time-dependent power laws while maintaining its shape in time~\cite{Berges:2015kfa}.
Near a nonthermal fixed point, the Betti number distributions of energy density sublevel sets can also reveal self-similarity~\cite{Spitz:2023wmn}.
In this work, the shape of the Betti number distributions shown in \Cref{fig:dim0Betti} is not globally preserved in time, in particular for dimension 0 at $N=1,2,3$, and dimension 1 at $N=2$. 
Yet, for these, the shape of the peaks at lower $\Delta T^{00}$-values remains approximately constant in time.
We discuss this in detail in \Cref{AppendixB} considering potential self-similar scaling. 

For clarity, here we focus on the time-dependence of the corresponding dimension-0 Betti number peak values, $\beta_{0,\max}^\topo(t) = \max\{\beta_0(t,\nu)\,|\,\nu\in \text{left peak}\}$.
This counts the maximum number of connected components formed by the pixelised sublevel sets as the filtration parameters corresponding to the peak are swept through, correlating with the number of defects.
If a power law in time with exponent $-\vartheta_N$ can be identified from $\beta_{0,\max}^\topo(t)$, i.e.,
\begin{equation}
\beta_{0,\max}^\topo(t) \sim t^{-\vartheta_N}, 
\end{equation}
then in three dimensional space, average length scales associated with the connected component configurations at the respective value of $\nu\equiv\Delta T^{00}$ grow as a power law with exponent $\vartheta_N/3$:
\begin{equation}
L_N(t) = \left(\frac{N_s^3 a_s^3}{\beta_{0,\max}^\topo(t)}\right)^{1/3}\sim t^{\vartheta_N/3}.
\end{equation}
Such a power law growth of length scales can be indicative of the self-similar scaling of corresponding Betti number distributions.
Since for $N=1,2,3$, the lower-$\Delta T^{00}$ peak is defect-related as noted in \Cref{sec:phdef}, this length scale can be sensitive to the dynamics of defects.
In particular, the number of strings correlating with the number of connected components of the $\Delta T^{00}$ sublevel sets (cf.~\Cref{defectindens}), $L_N(t)$ can serve as a proxy for the time-dependence of average length scales associated with string configurations.

In \Cref{fig:scaling}, we display the time-dependence of $\beta_{0,\max}^\topo(t)$ for (a) $N=1$, (b) $N=2$ and (c) $N=3$.
The main figures give results for averaging over every $8^3$ blocks, while the upper insets show results for $16^3$ blocks.
All $\beta_{0,\max}^\topo(t)$ decrease in time, which indicates that average length scales associated with the connected component configurations grow, as discussed previously (cf.~ \Cref{sec:phdef}).
The decrease of $\beta_{0,\max}^\topo(t)$ is of a power law form for long time ranges.
Fitting a power law to these curves for the indicated times%
\footnote{Note the different time intervals for $8^3$ versus $16^3$ blocks.
This is due to initially slower dynamics in Betti number distributions for averaging over $16^3$ blocks compared to $8^3$ blocks, since part of the faster ultraviolet modes do not contribute to the former (cf.~\Cref{fig:coarse}).} 
via standard $\chi^2$ fits yields $\vartheta_1 = 0.69\pm 0.02$~($N=1$), $\vartheta_2 = 0.64\pm 0.02$~($N=2$) and $\vartheta_3=1.19\pm 0.01$~($N=3$) when averaging over every $8^3$ blocks.
We obtain $\vartheta_1 = 0.67\pm 0.04$, $\vartheta_2 = 0.61 \pm 0.05$ and $\vartheta_3=1.23\pm 0.03$ when averaging over $16^3$ blocks.
Thus, within errors these exponents are insensitive to blockwise averaging in the considered regime.
They describe dynamics predominantly related to string defects, which appears well-separated from the dynamics on much smaller length scales based on the described insensitivity to blockwise averaging.
This is in contrast to the analogous analysis for the potentially defect-related peak at $N=2$ in dimension-1 Betti number distributions, for which the power law scaling behaviour is different when averaging over blocks of $8^3$ versus $16^3$ lattice points.
For this reason, a more detailed analysis of the defect-related peak in $N=2$ dimension-1 Betti number distributions is only described in \Cref{AppendixC}.

The fitted exponents $\vartheta_N$ correspond to the following power laws describing the time dependence of the length scales $L_N(t)$ (for averaging over every $8^3$ blocks):
\begin{equation}\label{EqLNResults}
    L_1(t)\sim t^{0.2},\quad L_2(t)\sim t^{0.2}, \quad L_3(t)\sim t^{0.4}.
\end{equation}

It is instructive to compare this with the known dynamics of topological defects as captured by phase-ordering kinetics~\cite{Bray:1994zz}, which describes the growth of order through coarsening dynamics when a system is quenched across a phase transition.
Notably, phase-ordering kinetics captures the mutual annihilation of vortices and anti-vortices (strings) as well as the shrinking of domain walls with time.
These processes can give rise to self-similar scaling which is characteristic of a nonthermal fixed point, with defect-related length scales displaying a power law in time.
For instance, numerical studies of universal dynamics in one-dimensional Bose gases have revealed power law exponents of order 0.25 to 0.35~\cite{Schmied:2018osf,  Fujimoto2019FlemishStringsMagneticSolitons, Siovitz:2023ius}.
Recent experiments with ultracold atoms in a quasi-one-dimensional elongated optical trap have pointed towards a growth of length scales related to vector solitons as a power law in time with exponent $0.28\pm 0.05$~\cite{Lannig:2023fzf}.
In two spatial dimensions it is known that inter-vortex distances can grow proportional to $t^{0.2}$ for nonrelativistic systems~\cite{Karl:2016wko, doi:10.1126/science.aat5793, Spitz:2020wej, wheeler2021relaxation}, and similarly for a nonrelativistic projection of relativistic scalar theory~\cite{Deng:2018xsk}.
For three spatial dimensions, signatures of defects in scalar field theories have been detected previously, albeit without analysing their self-similar scaling dynamics~\cite{Nowak:2010tm, Gasenzer:2011by, Nowak:2011sk, Berges:2017ldx}.
Yet, even in this case, phase-ordering kinetics predicts scaling exponents for defect-related length scales in the range of 0.2 to 0.3, if conserved order parameters are considered~\cite{Bray:1994zz}.
Our results for $L_1(t)$ and $L_2(t)$ as deduced with persistent homology are well within this range.
The scaling exponent of $L_3(t)$ is closer to $0.5$, which is considered to be the relevant exponent for length scale dynamics associated with the particle cascade in large-$N$ expansions (up to anomalous dimensions)~\cite{PineiroOrioli:2015cpb, Boguslavski:2019ecc, PineiroOrioli:2018hst}, where topological defects are absent.
Still, topological defects can also give rise to such dynamics~\cite{Bray:1994zz, Mikheev:2018adp}.

Note that the association of persistent homology classes with defects is not a one-to-one correspondence, and not all persistent homology classes need to behave uniformly according to~\eqref{EqLNResults}.
Therefore, for our method we expect systematic errors on the deduced scaling dynamics of topological defects, which we do not consider here and are to be discussed in a future work.

\subsection{Signatures of energy transport}\label{sec:energytransport}
In addition to the coarsening dynamics, there is an ongoing homogenisation process of local energy densities. 
This is apparent from the peaks with increasing amplitudes close to $\Delta T^{00} = 0$ in the Betti number distributions, which shifts towards the mean local energy density.
Again, we study their maxima, in this case for \mbox{dimension-1} Betti number distributions, denoted by $\beta_{1,\max}^\ener(t) = \max\{\beta_1(t,\nu)\,|\,\nu\in \text{right peak}\}$.

	\begin{figure}
		\includegraphics[scale=0.66]{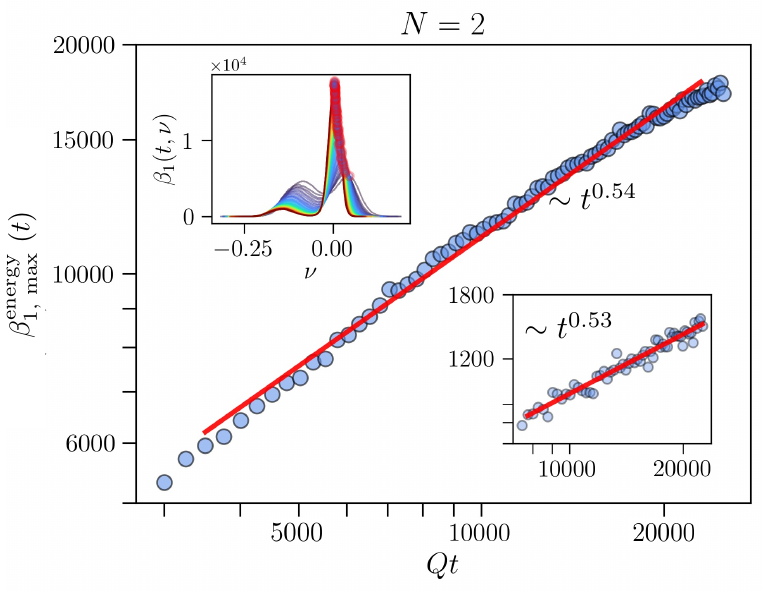}
		\caption{Temporal scaling of the right dimension-1 Betti number distribution peak values for $N=2$.
  Averaging over blocks of $8^3$ lattice sites has been employed.
  The top inset shows the actual Betti number distributions, where the red circles indicate the peak values.
  The bottom inset shows peak values for averaging over every $16^3$ blocks.
  The power law fit is based on the data in the $Qt$ range for which it is displayed.
  Note the smaller time interval shown here in comparison to \Cref{fig:scaling}.
  }
		\label{fig:scalingd12}
	\end{figure}

In \Cref{fig:scalingd12}, we show $\beta_{1,\max}^\ener(t)$ for $N=2$ when averaging over every $8^3$ blocks (main figure) and $16^3$ blocks (bottom inset).
Comparable outcomes can be obtained for other dimensions and $N$.
We find that the number of features as counted by $\beta_{1,\max}^\ener(t)$ in \Cref{fig:scalingd12} grows in time, such that corresponding dimension-1 holes (tunnels) steadily decrease in size.
Though slightly bent, the data can be approximately described by a power law, \mbox{$\beta_{1,\max}^\ener(t)\sim t^{-\vartheta_\ener}$}.
Standard $\chi^2$ fits yield $\vartheta_\ener = -0.54\pm 0.02$ for averaging over every $8^3$ blocks and $\vartheta_\ener = -0.53\pm 0.03$ for $16^3$ blocks, such that $\vartheta_\ener$ appears stable for this regime of blockwise averaging.

As before, we can estimate the time dependence of average length scales associated with the dimension-1 holes from the time dependence of $\beta_{1,\max}^\ener(t)$:
\begin{equation}
    L_\ener(t) = \left(\frac{N_s^3 a_s^3}{\beta_{1,\max}^\ener(t)}\right)^{1/3}\sim t^{\vartheta_\ener / 3},
\end{equation}
which leads to $L_\ener(t)\sim t^{-0.18}$ for averaging over every $8^3$ blocks.
It is well known that turbulent energy transport towards the ultraviolet is accompanied by self-similar scaling behaviour characteristic of a nonthermal fixed point, for which related length scales dynamically shrink as $\sim t^{-1/7}$~\cite{Micha:2002ey, Micha:2004bv}.
The power law behaviour of $L_\ener(t)$ is close to this, which suggests that the corresponding structures in Betti number distributions are due to excitations in local energy densities transported towards smaller length scales.
Deviations can be caused by the blockwise averaging procedure and the tentative interpretation of features in the Betti number distributions with different physical phenomena, which are in general not perfectly separated.
Moreover, while the maxima of the Betti number distributions can provide estimates for the overall number of structures associated with the corresponding peaks, this number partly remains ambiguous in light of non-uniform persistences of features in the filtration.

\section{Conclusions}\label{sec:discussion}
In this work, we have investigated the real-time dynamics of relativistic $\mathrm{O}(N)$ scalar fields in over-occupied scenarios. 
We have applied TDA to lattice configurations, which can provide complementary information to the traditionally investigated distribution functions computed from equal-time two-point correlations.
More specifically, we have considered Betti numbers computed for sublevel sets of local energy density fluctuations.
These have revealed clear signals of dynamically generated topological defects. 
The identification of defects in the Betti numbers is based on the comparison with defect structures visible in local energy density landscapes.
Crucially, the $N$-dependent topological features visible in the Betti numbers have been consistent with the classification of defects for condensates in relativistic $\mathrm{O}(N)$ scalar fields~\cite{Moore:2015adu}.

The number of connected components associated with the topological defects has decreased in time as a power law, which is indicative of self-similar scaling dynamics.
This behaviour corresponds to power law growth of length scales associated with their dilution, with scaling exponents $\sim 0.2$ for $N=1$ and $N=2$, and $\sim 0.4$ for $N=3$.
While the values for $N=1,2$ agree well with the findings for topological dynamics in a variety of simulations, the value for $N=3$ is closer to the scaling behaviour of the model at large~$N$.
Yet, topological dynamics can also lead to the latter scaling dynamics~\cite{Bray:1994zz, Mikheev:2018adp}.
A thorough quantitative analysis of the relation to phase-ordering kinetics and a more careful study of possible dynamical contributions to the structures associated with defects in Betti numbers is beyond the scope of this paper.

In addition, we have observed signatures of energy transport and the related universal scaling behaviour in the Betti numbers.
While for over-occupied scenarios the distribution function reveals a dual cascade with particle and energy transport, the Betti numbers computed for local energy density fluctuations may therefore be able to distinguish topological dynamics from energy transport.

Our work hints at the importance of topologically sensitive observables in order to uncover defect dynamics in universal regimes far from equilibrium.
In particular for three-dimensional systems, defect-driven temporal scaling dynamics can be hard to access with distribution functions and appears to be suppressed in over-occupied scenarios.
Based on the present work, TDA-based observables are particularly promising to investigate along with distribution functions, since they can be sensitive to extended field configurations of arbitrary size.

This method is applicable without major modifications to the analysis of ultracold atom experiments.
Local atom densities can play a similar role to the local energy densities considered in this work, with topological configurations also showing up as distinct minima.
These are accessible for instance using commonly employed absorption imaging techniques.
In this spirit, a recent work has exploited TDA to detect dark solitons in condensate density images~\cite{Leykam:2021xdt}.
It is particularly beneficial that this method does not rely on more sophisticated experimental techniques such as response measurements to probe correlation functions at unequal times~\cite{Chou:1984es, Uhrich2017NoninvasiveMeasurement, PineiroOrioli:2018hst, Boguslavski:2018beu, Boguslavski:2019ecc, Shen:2019jhl}.

Our work reinforces the potential of TDA to provide relevant information on the dynamics of strongly correlated many-body systems.
The choice of cubical complexes has especially been advantageous to reveal the presence of non-local structures.
TDA can facilitate the systematic study of the role of topological defects for nonequilibrium quantum many-body dynamics, with diverse regimes of applicability ranging from cold atoms to the collisions of heavy nuclei.

\begin{acknowledgments}
We thank J.~Berges, K.~Boguslavski, T.~Gasenzer, A.~N.~Mikheev and J.~M.~Pawlowski for fruitful discussions, furthermore, K.~Boguslavski and T.~Gasenzer for proof-reading the manuscript and K.~Boguslavski for his simulation code.
We thank D.~Kirchhoff for the collaboration on related work.
The authors acknowledge support by the state of Baden-Württemberg through bwHPC. 
This work is part of and funded by the Deutsche Forschungsgemeinschaft (DFG, German Research Foundation) under Germany’s Excellence Strategy EXC 2181/1–390900948 (the Heidelberg STRUCTURES Excellence Cluster) and the Collaborative Research Centre, Project-ID No. 273811115, SFB 1225 ISOQUANT. 
\end{acknowledgments}

	
\appendix

\section{Dependence of Betti numbers on the renormalised mass value}\label{AppendixMassDiscussion}
The simulations reported about in the main text have used a comparably large renormalised mass with $M=2Q$.
In this appendix, we discuss the dependence of the persistent homology results provided in \Cref{sec:phdef} on this choice.

\begin{figure}[!t]
	\includegraphics[scale=0.8]{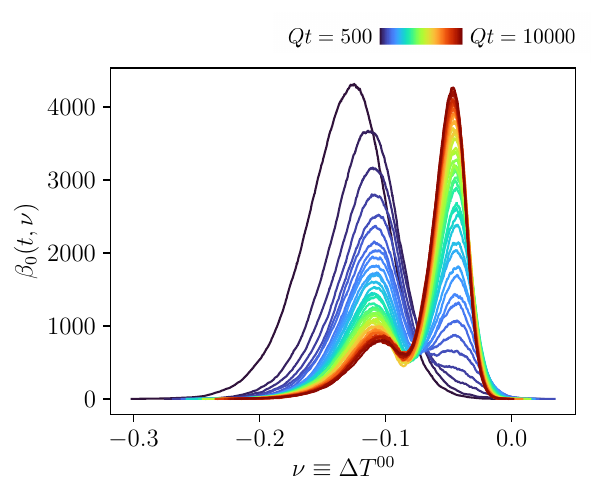}
	\caption{Dimension-0 Betti numbers for $\Delta T^{00}$ sublevel sets for $N=2$ field components, with averaging over blocks of $8^3$ lattice sites. 
 The renormalised mass used to generate these results is $M^2 = Q^2/4$.}\label{FigBetti0SmallerMass}
\end{figure}

\Cref{FigBetti0SmallerMass} provides the dimension-0 Betti numbers for simulations with $N=2$ field components and renormalised mass $M =Q/2$.
Comparison with \Cref{fig:dim0Betti}, which has given the corresponding results for $M=2Q$, reveals that the right peak close to $\nu\equiv \Delta T^{00}=0$ increases in height for the smaller mass value, while the height of the left peak appears roughly insensitive to the choice of mass. 
In particular, the scaling behaviour of $\beta_{0,\max}^{\mathrm{topo}}(t)$ remains the same compared to the larger mass (not shown).

We can heuristically understand this behaviour as follows.
For the results displayed in \Cref{fig:dim0Betti} and \Cref{FigBetti0SmallerMass}, we have employed averaging over blocks of $8^3$ lattice sites.
Removing ultraviolet fluctuations, this emphasises structures in the infrared, where the model is well-described by a nonrelativistic complex scalar field theory~\cite{PineiroOrioli:2015cpb, Deng:2018xsk}.
The Hamiltonian of such a theory comes with a kinetic term $\sim \mathbf{k}^2/2M$, where $\mathbf{k}$ denotes the spatial momentum of a Fourier mode of the complex-valued field.
A larger value for $M$ therefore suppresses the contributions of spatial gradients of local fluctuations to energy densities, which pronounces structures with small spatial gradients, for instance topological defects.

Moreover, we have argued in favour of the association of the right peak in the dimension-0 Betti numbers with energy transport towards the ultraviolet, see \Cref{sec:energytransport}.
In the $\Delta T^{00}$ snapshots presented in \Cref{fig:energies1}, this corresponds to the local fluctuations on small length scales.
A larger mass therefore suppresses these, such that the topological defects appear pronounced relative to the small-scale fluctuations, even though their number might remain roughly invariant under the choice of mass.
This explains the behaviour of the dimension-0 Betti numbers shown in \Cref{FigBetti0SmallerMass}.

For completeness, we note that the occupation number distributions computed for both mass values agree (not shown) and are similar to the results of~\cite{PineiroOrioli:2015cpb}.

\section{The classification of topological defects in condensates of the $\mathrm{O}(N)$ vector model} \label{AppendixA}
In this appendix, we derive the classification of topological defects in condensates of the $\textrm{O}(N)$ vector model in detail.
The arguments are drawn from~\cite{Moore:2015adu} and expanded here.

Intuitively, topological defects form when growing occupancies in the infrared organise into condensates along with the inverse particle cascade.
To this end, the behaviour of spatial zero modes of the field variables is of relevance for the classification of topological defects.
In particular, it is the topology of the phase space of zero modes of the field variables, $(\tilde{\phi}_a,\partial_t \tilde{\phi}_a)_{a=1,\dots,N}$, where $\tilde{\phi}_a \equiv \tilde{\phi}_a(t)\equiv \tilde{\phi}_a(t,\textbf{p}=0)$, along with dynamical constraints which determines the latter.
Denoting this space for the $\textrm{O}(N)$ vector model by $\mathscr{C}_N$, we first motivate its general construction from first principles based on the condensate dynamics.
We subsequently describe the specific structure of the $\mathscr{C}_N$ and their topological properties, ultimately leading to the classification of defects in three spatial dimensions.

\subsection{Dynamically realised condensate phase space}
In our simulations, the modes in the infrared are highly occupied at early times for the considered initial conditions and are thus well-described classically.
Here, we derive an approximate equation of motion for the condensate.
First, the classical inverse propagator follows from the action \eqref{eq:action} as
\begin{align}
    &-i\frac{\delta^2 S[\phi]}{\delta \phi_a(x) \delta\phi_b(y)}\nonumber\\
    &\quad = i\delta(x-y) \bigg[\delta_{ab}(\square + m^2)  + \frac{\lambda}{6N} \bigg(2\phi_a(x) \phi_b(x)\nonumber\\
    &\qquad\qquad\qquad\qquad\qquad\qquad + \delta_{ab} \sum_c \phi_c(x)^2\bigg)\bigg]
\end{align}
with $\square \equiv \partial_\mu \partial^\mu$ and where we explicitly denoted the summation over the field components.
This leads to the classical self-energy matrix
\begin{equation}\label{eq:selfenergies}
    \Sigma_{ab}^\cl(x) = \frac{\lambda}{6N}\bigg(2 \phi_a(x) \phi_b(x)  + \delta_{ab} \sum_{c=1}^N  \phi_c(x)^2 \bigg).
\end{equation}
Its spatial Fourier-transform is $\tilde{\Sigma}^\cl_{ab}(t,\pp)=\int _\xx \Sigma_{ab}^\cl(t,\xx)\exp(-i\pp\xx)$.
The action \eqref{eq:action} yields an inhomogeneous Klein-Gordon equation as the classical equation of motion for spatial zero modes of the fields:
\begin{equation}\label{eq:effkleingordon}
    \sum_{b=1}^N \bigg[\delta_{ab} (\partial_t^2 + m^2)\tilde{\phi}_b(t) + \int_{\pp}\tilde{\Sigma}_{ab}^\cl(t,-\pp) \tilde{\phi}_b(t,\pp) \bigg] = 0,
\end{equation}
where $\int_\pp\equiv \int \dd ^3 \pp / (2\pi)^3$ and the dependence on the classical self-energy matrix $\tilde{\Sigma}^\cl_{ab}(t,-\pp)$ has been explicitly denoted.
We assume a large fraction of the particles occupies the zero mode, so that the momentum integral in \Cref{eq:effkleingordon} is dominated by momentum zero.
This finally yields the equation
\begin{equation}\label{eq:effkleingordonzeromodes}
    \sum_{b=1}^N [\delta_{ab} (\partial_t^2 + m^2) + \tilde{\Sigma}^\cl_{ab}(t,\pp=0)]\tilde{\phi}_b(t) = 0,
\end{equation}
which governs the time evolution of the condensate.

For a single field component ($N=1$) \mbox{$\tilde{\phi} \equiv \tilde{\phi}_{a=1}$} and the effective mass squared is \mbox{$M^2 = m^2 + \tilde{\Sigma}_{11}^\cl$}.
\Cref{eq:effkleingordon} leads to the decomposition $\tilde{\phi}=\tilde{\phi}_0 \cos(M t+\delta)$, where $\tilde{\phi}_0$ is a real-valued peak amplitude of the field, $\delta$ a phase and $M$ determines the oscillation frequency of the field.
The amplitude $|\tilde{\phi}_0|$ is fixed by transient particle number conservation.%
\footnote{This is similar to nonrelativistic scalar fields for which particle number conservation is exact.}
From a topological viewpoint, the $N=1$ condensate phase space $\mathscr{C}_1$ is thus described by a circle, $\mathscr{C}_1\simeq S^1$.

For $N\geq 2$ field components the solutions \mbox{$\tilde{\phi}\equiv(\tilde{\phi})_{a=1,\dots,N}$} to \Cref{eq:effkleingordonzeromodes} generally have the form of oscillations sweeping out ellipses in the internal space of field components.
These can interpolate between the extreme possibilities of $\tilde{\phi}$ and $\partial_t \tilde{\phi}$ parallel, so that oscillations happen along a line in this space, or $\tilde{\phi}$ and $\partial_t \tilde{\phi}$ orthogonal, so that oscillations happen along a circular orbit.
It can be shown, that the energy-minimising condensate configurations are not straight lines and, therefore, have the topology of circular orbits.
Indeed, repeating the argument provided in~\cite{Moore:2015adu}, we can understand this by comparing the energy cost of a circular orbit to that of a straight line in $\mathrm{O}(N)$ field space.
The energy density $\varepsilon$ of the zero mode of a single-component scalar field oscillating back and forth in a quartic potential is
\begin{equation}\label{EqQuarticPotentialN1}
    \varepsilon=\frac{\lambda}{8} \tilde{\phi}_0^4=\frac{1}{2} (\partial_t\tilde{\phi})^2+\frac{\lambda}{8} \tilde{\phi}^4,
\end{equation}
where we have neglected the mass term. 
This equation can be rearranged for $\partial_t\tilde{\phi}$ to obtain a periodic solution, which has oscillation frequency
\begin{equation}
    \omega^2 = \frac{\pi \Gamma(3 / 4)^2}{\Gamma(1 / 4)^2} \lambda \tilde{\phi}_0^2.
\end{equation}
The particle number stored in the condensate is $n(\varepsilon)=\int_0^{\varepsilon} \dd \varepsilon^{\prime}/\omega = 4\varepsilon/(3\omega)$, and using \Cref{EqQuarticPotentialN1}, we obtain 
\begin{equation}
    \varepsilon = (8\pi^2)^{1/3}\left[\frac{3\Gamma(3/4)}{4\Gamma(1/4)}\right]^{4/3}\lambda^{1 / 3} n^{4 / 3}.
\end{equation}
On the other hand, for a scalar field with two or more components with a circular orbital motion in the same potential, the energy density is
\begin{equation}
    \varepsilon = \frac{1}{2}(\partial_t\tilde{\phi}_0)^2 +\frac{\lambda}{8}\tilde{\phi}_0^4 = \frac{3}{4} (\partial_t\tilde{\phi}_0)^2 = \frac{3\omega^2}{4}\tilde{\phi}_0^2 = \frac{3\lambda}{8}\tilde{\phi}_0^4,
\end{equation}
where we have used the Virial theorem $(\partial_t\tilde{\phi})^2/2=\lambda\tilde{\phi}_0^4/4$ and $\tilde{\phi}_0=\tilde{\phi}(t_0)$ denotes the condensate configuration at some initial time $t_0$.
Hence, 
\begin{equation}
    \omega^2=\frac{\lambda}{2}\tilde{\phi}_0^2=\frac{2^{1/2}}{3^{1/2}}\lambda^{1/2}\varepsilon^{1/2}.
\end{equation}
Using the same arguments as before for $n(\varepsilon)$, we get that 
\begin{equation}
    \varepsilon = \frac{3}{2^{7/3}}\lambda^{1/3}n^{4/3}.
\end{equation}
Therefore, comparing the numerical prefactors, at fixed particle number the energy cost of a circular orbit is lower than that of a straight line.
We can conclude that energy-minimising orbits $\tilde{\phi}(t)$ are not straight lines and thus homotopy-equivalent to circles in the zero mode phase space.
We expect that the addition of a non-zero mass term does not alter the topology of energy-minimising orbits.
This is \emph{a posteriori} reinforced by the consistency of our work with the defect classification outlined here, which relies on orbits being homotopy-equivalent to circles.

We thus proceed with the consideration of circular orbits in the zero mode phase space.
For these the constancy of $|\tilde{\phi}|$ due to effective particle number conservation is complemented by $|\partial_t \tilde{\phi}|$ being constant due to the circular orbit geometry, along with the orthogonality constraint $\tilde{\phi}\perp \partial_t\tilde{\phi}$ in the internal field space.
In general, a fixed-length $\tilde{\phi}$ is an element of the $(N-1)$-sphere $S^{N-1}\subset \rr^N$.
Any tangent vector to $S^{N-1}$ at a point $\tilde{\phi}\in S^{N-1}$ is orthogonal to the vector $\tilde{\phi}$ itself within $\mathbb{R}^N$.
The vectors $\partial_t\tilde{\phi}$ orthogonal to $\tilde{\phi}$ actually form the tangent manifold $T_{\tilde{\phi}}S^{N-1}$.
The constancy constraint for $|\partial_t \tilde{\phi}|$ singles out tangent vectors of constant length.
Since topology does not discriminate between the length of such tangent vectors, the constancy constraint for $|\partial_t \tilde{\phi}|$ can be taken care of upon restricting to normalised tangent vectors $\partial_t \tilde{\phi}$ with $|\partial_t \tilde{\phi}|=1$.%
\footnote{For the same reason we here ignored the mass dimension of $|\partial_t\tilde{\phi}|$.}
Finally, for $N\geq 2$, the condensate phase space $\mathscr{C}_N$ is (homotopy-equivalent to) the unit tangent bundle of $S^{N-1}$, $\mathscr{C}_N \simeq T^1 S^{N-1}$, which as a set reads
\begin{equation}
    T^1 S^{N-1} = \bigcup_{\tilde{\phi}\in S^{N-1}} \{(\tilde{\phi}, \partial_t\tilde{\phi})\,|\, \partial_t\tilde{\phi}\in T_{\tilde{\phi}}S^{N-1},\; |\partial_t\tilde{\phi}|=1\}.
\end{equation}

\subsection{Homotopy groups of $\mathscr{C}_N$}
The topology of $\mathscr{C}_N$ can be nontrivial as quantified by the low-order homotopy groups, which is the mathematical origin of topological defects.
Specifically, a nontrivial zeroth homotopy group $\pi_0(\mathscr{C}_N)$ indicates that $\mathscr{C}_N$ comprises different connected components, i.e., the condensate is separated into different domains bounded by domain walls.
If the fundamental group $\pi_1(\mathscr{C}_N)$ is nontrivial, $\mathscr{C}_N$ is not simply connected, and string defects can occur, i.e., vortex lines.
If the second homotopy group $\pi_2(\mathscr{C}_N)$ is nontrivial, then monopole defects can occur.
A nontrivial third homotopy group $\pi_3(\mathscr{C}_N)$ would indicate the possibility of textures.
However, these have no defect core and are unstable~\cite{Turok:1989ai}, such that we exclude them from our analysis.
Apart from textures, strings, domain walls and monopoles are all types of defects which can occur in three spatial dimensions.
Investigating the presence of topological defects thus requires the computation of the homotopy groups $\pi_\ell(\mathscr{C}_N)$ for $\ell=0,1,2$ and different $N$.

\subsubsection{$N=1$}
For $N=1$, the condensate phase space is $\mathscr{C}_1\simeq S^1$, for which $\pi_1(S^1)\cong \zz$ and all other homotopy groups vanish, such that only string defects can occur.

\subsubsection{$N=2$}
For $N=2$, we have to consider $\mathscr{C}_2\simeq T^1S^1$.
The unit tangent line at any point in $S^1$ consists of exactly two points, such that $\mathscr{C}_2\simeq S^1\times \zz_2$.
The homotopy groups of such products factorise~\cite{HatcherAlgebraicTopology}, such that $\pi_\ell(\mathscr{C}_2) \cong \pi_\ell(S^1)\times \pi_\ell(\zz_2)$ for all $\ell\in \nn$.
We find the nontrivial homotopy groups $\pi_0(\mathscr{C}_2)\cong \pi_0(S^1)\times \pi_0(\zz_2) \cong 0\times \zz_2\cong \zz_2$ and $\pi_1(\mathscr{C}_2)\cong \pi_1(S^1)\times \pi_1(\zz_2)\cong \zz\times 0\cong \zz$ and all other homotopy groups vanish.
Hence, both domain walls and string defects can occur.

\subsubsection{$N=3$}
For $N=3$, we have to consider $\mathscr{C}_3\simeq T^1 S^2$.
This space can be identified with $\textrm{SO}(3)$, since $\tilde{\phi}$ describes a direction in $\rr^3$ and $\partial_t \tilde{\phi}$ is orthogonal to it. 
Together with their cross product, they single out an orthonormal coordinate frame.
The fundamental group is nontrivial: $\pi_1(\mathscr{C}_3)\cong \pi_1(\textrm{SO}(3))\cong  \zz_2$ and all other homotopy groups vanish.
There are thus string defects.

\subsubsection{$N=4$}
For $N=4$, we have to consider $\mathscr{C}_4\simeq T^1 S^3$.
The 3-sphere is parallelizable, i.e., $TS^3 \cong S^3\times \rr^3$ is a trivial bundle above $S^3$.
The unit tangent vector constraint then singles out $\mathscr{C}_4\simeq S^3\times S^2$.
Again, the homotopy groups of such products factorise~\cite{HatcherAlgebraicTopology}, such that $\pi_\ell(\mathscr{C}_4)\cong \pi_\ell(S^3)\times \pi_\ell(S^2)$ for all $\ell$.
With the homotopy groups of the spheres we find that $\pi_2(\mathscr{C}_4)\cong \zz$, $\pi_3(\mathscr{C}_4)\cong \zz^2$ and all other homotopy groups vanish.
While the former ($\pi_2$) indicates that there are monopole defects, the latter ($\pi_3$) is irrelevant for defects in three spatial dimensions as considered here.

\subsubsection{$N\geq 5$}
For general $N\geq 5$ one needs to consider $\mathscr{C}_N\simeq T^1(S^{N-1}) \cong \mathrm{Spin}(N)/\mathrm{Spin}(N-2)$.
The identification with the spin group quotient comes about since the elements of the unit tangent bundle of $S^{N-1}$ single out orthonormal 2-frames in $\rr^N$, which form the second Stiefel manifold $V_2(\rr^N)$.
Specifically, for $N\geq 3$ we have that $V_2(\rr^N)$ is diffeomorphic to \mbox{$\mathrm{SO}(N)/\mathrm{SO}(N-2)$}~\cite{HatcherAlgebraicTopology}.
This in turn is diffeomorphic to the quotient $\mathrm{Spin}(N)/\mathrm{Spin}(N-2)$ for all $N\geq 3$, as can be seen with standard results for homogeneous spaces together with the consideration of relevant stabiliser subgroups of $\mathrm{SO}(N)$ and $\mathrm{Spin}(N)$.

The computation of the homotopy groups of this space requires more advanced methods from algebraic topology~\cite{HatcherAlgebraicTopology}, which we only briefly outline here.
Homotopy groups of the quotients $\mathrm{Spin}(N)/\mathrm{Spin}(N-2)$ can be computed with the long exact sequence of relative homotopy groups, which reduces their computation to the homotopy groups of the spin groups themselves.
The spin groups are the universal covers of the special orthogonal groups for $N\geq 3$, as indicated by the related short exact sequence of groups:
\begin{equation}\label{eq:spingroupexactseq}
    1\to \zz_2\to\mathrm{Spin}(N)\to \mathrm{SO}(N)\to 1.
\end{equation}
The spin groups are thus simply connected, so that $\pi_0(\mathrm{Spin}(N))\cong \pi_1(\mathrm{Spin}(N))\cong 0$ for all \mbox{$N\geq 3$}.
Moreover, the short exact sequence \eqref{eq:spingroupexactseq} induces a long exact sequence of homotopy groups.
Using \mbox{$\pi_2(\mathrm{SO}(N))\cong 0$} for all $N\geq 2$ (in the stable range), we then find $\pi_2(\mathrm{Spin}(N))\cong 0$ for all $N\geq 2$.
Together with the previously mentioned long exact sequence of relative homotopy groups this yields $\pi_0(\mathscr{C}_N)\cong 0$, $\pi_1(\mathscr{C}_N)\cong 0$ and $\pi_2(\mathscr{C}_N)\cong 0$ for all $N\geq 5$, while higher homotopy groups can be nontrivial.
Yet, it is these homotopy groups which determine the topological defects for three spatial dimensions.
Topological defects are thus absent in the considered three-dimensional $\textrm{O}(N)$ vector model for all $N\geq 5$.

\section{Details on the lower star filtration of cubical complexes and persistent homology}\label{AppendixMathDetailsComplexesPersHom}
This appendix provides mathematical details on the construction of the lower star filtration of cubical complexes, homology groups and persistent homology.
It closely follows similar expositions in~\cite{Spitz:2023wmn}.

\subsection{Lower star filtration}
We introduce the lower star filtration for a real-valued lattice function $f:\Lambda_s \to \rr$ such as $\Delta T^{00}(t,\cdot)$, intuitively corresponding to a pixelisation of its lattice sublevel sets.
To begin with, let $\mathcal{C}$ denote the full cubical complex of the lattice $\Lambda_s$, consisting of one 3-cube $\xx + [-a_s/2,a_s/2]^3$ for each spatial lattice point $\xx\in\Lambda_s$. 
$\mathcal{C}$ also includes all faces, edges and vertices of every 3-cube, such that it is closed under taking boundaries.
$\mathcal{C}$ is equipped with the information contained in the function $f$ by means of inductively constructing a map $F:\mathcal{C}\to \rr$. 
By construction, any 3-cube $C\in \mathcal{C}$ has a unique lattice point $\xx\in\Lambda_s$ at its center, so that we can set $F(C):=f(\xx)$.
For all 2-cubes $D\in\mathcal{C}$ we set
\begin{equation}\label{EqInductiveDefTt}
F(D):=\min\{F(C)\,|\, D\subset \partial C,\, C\in\mathcal{C}\text{ 3-cube}\}.
\end{equation}
\Cref{EqInductiveDefTt} is then applied inductively to construct $F$ for lower-dimensional cubes from higher-dimensional ones, until $F$ is defined on all $\mathcal{C}$. 
This construction is called the lower star filtration.

We define cubical complexes corresponding to lattice sublevel sets of $f$ as
\begin{equation}
\mathcal{C}_{f}(\nu):=F^{-1}(-\infty,\nu].
\end{equation}
Indeed, these are closed under taking boundaries.
As stated in the main text, they form a filtration: whenever $\nu\leq\mu$, we have the inclusion $\mathcal{C}_{f}(\nu)\subseteq \mathcal{C}_{f}(\mu)$.

\subsection{Homology groups}
Let $\mathcal{C}$ be a cubical complex, although the construction of homology groups is the same as for simplicial complexes.
More details can be found e.g.~in~\cite{otter2017roadmap} and references cited therein.
In this work, we focus on chain complexes and homology groups with $\zz_2$-coefficients, such that the $k$-th chain complex $C_k(\mathcal{C})$ of $\mathcal{C}$ consists of formal sums of chains of $k$-cubes with $\zz_2$-coefficients. 
The boundary operator ${\partial_k:C_k(\mathcal{C})\to C_{k-1}(\mathcal{C})}$ is defined to map a chain of $k$-cubes to its boundary, which is a $(k-1)$-chain. 
Since boundaries of such chain boundaries are empty, $\partial_{k-1}\circ \partial_k = 0$. 
We define the cycle group as $Z_k(\mathcal{C}):=\ker (\partial_k)$, which consists of all closed $k$-chains, i.e., $k$-chains without boundary.
The boundary group can be defined as $B_k(\mathcal{C}):=\mathrm{im}(\partial_{k+1})$, consisting of all those $k$-chains, which are boundaries of $(k+1)$-chains. 
As subgroups, $B_k(\mathcal{C})\subseteq Z_k(\mathcal{C})$, such that their quotient groups are well-defined:
\begin{equation}
H_k(\mathcal{C}):= Z_k(\mathcal{C})/B_k(\mathcal{C}).
\end{equation}
These are called homology groups.

The topology of $\mathcal{C}$ can be studied via the homology groups $H_k(\mathcal{C})$.
They capture similar topological information compared to homotopy groups of $\mathcal{C}$, but are often not the same. 
Elements of $H_k(\mathcal{C})$ are called homology classes and form equivalence classes of $k$-cycles, defined modulo higher-dimensional boundary contributions. 
Intuitively, they can be thought of as independent holes. 
Their number is given by the $\zz_2$-dimension of $H_k(\mathcal{C})$:
\begin{equation}
\beta_k(\mathcal{C}) := \dim_{\zz_2}(H_k(\mathcal{C})),
\end{equation}
which is called the $k$-th Betti number.

\subsection{Persistent homology groups}
Let $\{\mathcal{C}_\nu\}_{\nu\in \rr}$ be a filtration of complexes, such as $\mathcal{C}_\nu = \mathcal{C}_f(\nu)$ for the energy density sublevel set filtration considered in this work. 
Suppose we compute all their individual homology groups $\{H_k(\mathcal{C}_\nu)\}_\nu$. 
In addition, the filtration contains for all $\nu \leq \mu$ the inclusion maps $\mathcal{C}_\nu\hookrightarrow \mathcal{C}_\mu$, which induce maps on the homology groups:
\begin{equation}
\iota_k^{\nu,\mu}: H_k(\mathcal{C}_\nu)\to H_k(\mathcal{C}_\mu).
\end{equation}
The $\iota_k^{\nu,\mu}$ map a homology class in $H_k(\mathcal{C}_\nu)$ either to one in $H_k(\mathcal{C}_\mu)$, if it is still present for $\mathcal{C}_\mu$, or to zero, if corresponding (potentially deformed) cycles appear as boundaries in $H_k(\mathcal{C}_\mu)$. 
Moreover, nontrivial cokernels can appear for $\iota_k^{\nu,\mu}$: new homology classes can appear in $\mathcal{C}_\mu$, which are not in $\mathcal{C}_\nu$. 
The parameter $\mu$ can be chosen, so that for sufficiently small $\epsilon > 0$:
\begin{equation}\label{EqHomologyClassBirth}
H_k(\mathcal{C}_{\mu-\epsilon}) \subsetneq H_k(\mathcal{C}_\mu).
\end{equation}
The collection $\{(H_k(\mathcal{C}_\nu),\iota_k^{\nu,\mu})\}_{\nu\leq \mu}$ forms a so-called persistence module, which is tame, if \eqref{EqHomologyClassBirth} holds only for finitely many distinct values of $\mu$. 

By the structure theorem of persistent homology (see e.g.~\cite{otter2017roadmap} and references therein), any tame persistence module is isomorphic to its persistence diagram, i.e., the collection of all the birth-death pairs $(b,d)$, \mbox{$b<d\in \rr\cup\{\infty\}$}. 
Persistence diagrams are multisets, so the same birth-death pair may appear multiple times.

	\begin{figure}[!htb]
	\includegraphics[scale=0.62]{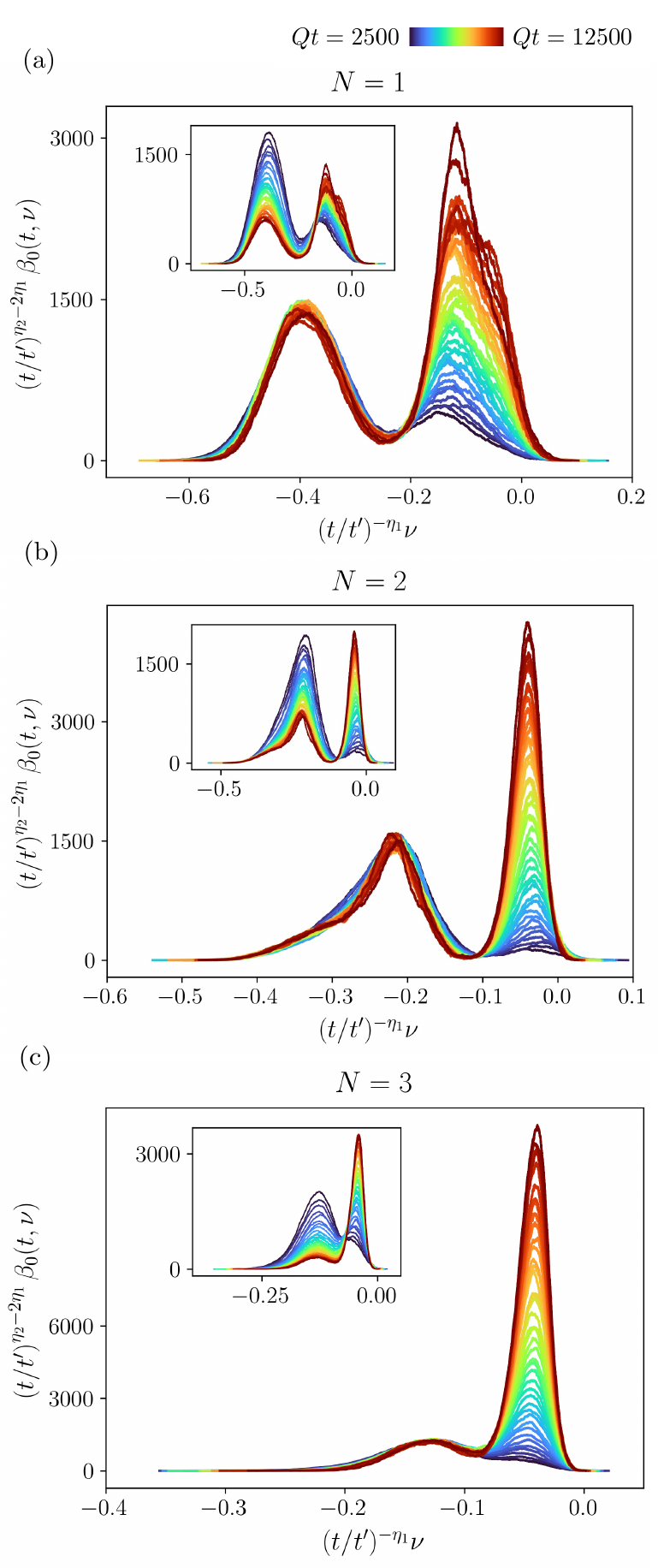}
	\caption{Rescaled dimension-0 Betti number distributions for (a) $N=1$, $\eta_1=0.02$, $\eta_2=0.73$, (b) $N=2$, $\eta_1=0.02$, $\eta_2=0.65$, (c) $N=3$, $\eta_1=0.01$, $\eta_2=1.20$. Blocks of $8^3$ lattice points have been averaged. The insets show the non-rescaled distributions.}\label{FigBettiRescaled}
\end{figure}

\section{Self-similar scaling in persistent homology} \label{AppendixB}

In this appendix, we discuss the self-similarity of Betti number distributions as introduced in~\cite{Spitz:2020wej} and similarly employed in~\cite{Spitz:2023wmn}.
For this, we need to introduce the so-called persistence pair distribution.
The persistent homology of the energy density filtration is fully described by the persistence diagram, which consists of all birth-death pairs $(b, d)$.
We denote it for dimension-$\ell$ features as $\mathrm{Dgm}_{\ell,i}(t)$, computed for a classical-statistical realisation $\phi_i(t,\xx)$.
The dimension-$\ell$ persistence pair distribution is then given by~\cite{Spitz:2020wej}
	\begin{equation}
		\mathfrak{P}_{\ell,i}(t, b, d)=\sum_{(b', d') \in \mathrm{Dgm}_{\ell,i}(t)} \delta(b-b') \delta(d-d').
	\end{equation}
Its expectation value can exist and is in general no longer a sum of Dirac $\delta$-functions anymore~\cite{spitz2020self}. 
In particular, it scales self-similarly in time~\cite{Spitz:2020wej} if 
	\begin{align}
			& \langle\mathfrak{P}_\ell\rangle(t, b, d) \nonumber\\
			& \quad=(t / t')^{-\eta_2}\langle\mathfrak{P}_\ell \rangle (t^{\prime},(t / t')^{-\eta_1} b,(t / t')^{-\eta_1} d), 
		\label{eq:ss}
	\end{align}
where $t,t'$ is any pair of times in the temporal regime of self-similar scaling and $\eta_1,\eta_2$ are suitable scaling exponents.

From $\langle\mathfrak{P}_\ell\rangle(t, b, d)$, the Betti number distribution can be computed as 
	\begin{equation}
		\langle \beta_\ell\rangle(t, \nu)= \int_{-\infty}^\nu \mathrm{d} b \int_\nu^{\infty} \mathrm{d} d \, \langle\mathfrak{P}_l\rangle(t, b, d).
	\end{equation}
If $\langle\mathfrak{P}_\ell\rangle(t, b, d)$ scales self-similarly in time, the Betti number distributions $\langle \beta_\ell\rangle(t,\nu)$ fulfil
	\begin{equation}\label{eq:ssb}
		\langle\beta_\ell\rangle(t, \nu) = (t / t')^{2\eta_1- \eta_2}\langle\beta_\ell \rangle (t',(t / t')^{-\eta_1} \nu).
	\end{equation}

In \Cref{FigBettiRescaled}, we show dimension-0 Betti number distributions for (a) $N=1$, (b) $N=2$ and (c) $N=3$, rescaled in time according to \Cref{eq:ssb} (averaging over every $8^3$ blocks).
The optimal scaling exponents $\eta_1$ and $\eta_2$ for the data to match the scaling behaviour \eqref{eq:ssb} are extracted using the self-similarity fitting protocol detailed in~\cite{PineiroOrioli:2015cpb}.
The fits take all times in the interval from $Qt=2500$ to $Qt=12500$ into account.
They are done for those filtration parameter ranges which correspond to the defect-related peaks in Betti number distributions: for $N=1$ from $\nu=-0.45$ to $-0.29$, for $N=2$ from $\nu=-0.31$ to $-0.15$, and for $N=3$ from $\nu=-0.20$ to $-0.09$. 
This way, the optimised scaling exponents are
\begin{equation}\label{eq:scalingexponentsN1}
    \eta_1 = 0.02 \pm 0.03,\qquad \eta_2 = 0.72\pm 0.01
\end{equation}
for $N=1$,
\begin{equation}\label{eq:scalingexponentsN2}
    \eta_1 = 0.01\pm 0.02,\qquad \eta_2 = 0.69\pm 0.03
\end{equation}
for $N=2$, and 
\begin{equation}\label{eq:scalingexponentsN3}
    \eta_1 = 0.01\pm 0.02,\qquad \eta_2 = 1.20\pm 0.01
\end{equation}
for $N=3$.
Indeed, rescaling the Betti number distributions with these scaling exponents consistently leads at least to approximate constancy in time, see \Cref{FigBettiRescaled}, in particular for $N=1$ and $N=3$.
Yet, smaller systematic deviations remain, since the shape of the peaks is not fully independent of time.

According to the Betti number scaling \eqref{eq:ssb}, the peak Betti number scales $\sim t^{2\eta_1 - \eta_2}$.
The numbers provided by the self-similarity fits, \Cref{eq:scalingexponentsN1,eq:scalingexponentsN2,eq:scalingexponentsN3}, match the analysis results provided in \Cref{sec:scaling}.
There, we also discuss the value of the exponent $2\eta_1-\eta_2$ in light of phase-ordering kinetics.

The value of $\eta_1$ is consistent with zero within errors.
If connected components at these $\Delta T^{00}$ values are primarily due to defects, this can indicate that defects appear locally at particular \emph{constant} energy density values.
We note that a zero result for $\eta_1$ is not in line with the packing relation~\cite{spitz2020self}, which yields $\eta_2 = 3\eta_1$ for energy conservation, providing a one-dimensional constraint on the filtration.
This is not in contradiction with~\cite{spitz2020self}, since the packing relation has been proven for self-similar scaling that applies to the entire filtration range.
This is not the case here, where the ongoing homogenisation of small-scale structures in energy densities provides features in Betti number distributions which cannot be rescaled with the exponents of \eqref{eq:scalingexponentsN1} to \eqref{eq:scalingexponentsN3} (cf.~the right peaks in \Cref{FigBettiRescaled}).

	\begin{figure}
		\includegraphics[scale=0.66]{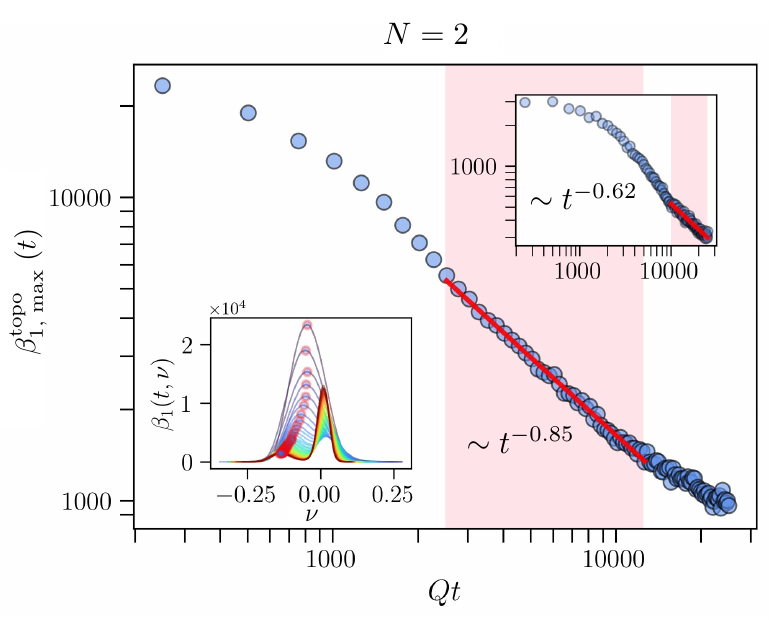}
		\caption{Temporal scaling of the left dimension-1 Betti number distribution peak values for $N=3$.
  Averaging over blocks of $8^3$ lattice sites has been employed.
  The bottom inset shows the actual Betti number distributions, where the red circles indicate the peak values. 
  The top inset shows peak values for averaging over every $16^3$ blocks.
  The power law fits are based on the data in the red shaded $Qt$ ranges. 
  }
		\label{fig:scalingd1}
	\end{figure}

\section{Maxima of dimension-1 Betti number distributions at $N=2$}\label{AppendixC}

This appendix discusses the defect-related left peak in the dimension-1 Betti number distributions for $N=2$ (cf.~\Cref{fig:dim0Betti}).
In \Cref{fig:scalingd1} we show maximum values of the peak depending on time.
The main figure has been computed for averaging over every $8^3$ blocks, the top inset for averaging over $16^3$ blocks.
As it is clearly visible, the numbers of dimension-1 features decline with time, such that the average distances associated with the structures grow.
Qualitatively, the discussion of \Cref{sec:scaling} applies here in an analogous way, so that we can associate this to the coarsening dynamics, potentially for domain walls.

Again, for an intermediate time range a power law can be fitted.
For averaging over every $8^3$ blocks, the peak values decrease as a power law with exponent $-0.85\pm 0.01$, and with exponent $-0.62\pm 0.03$ for averaging over $16^3$ blocks.
While the corresponding length scales of dimension-1 features thus grow as power laws with exponents well within the regime expected for coarsening dynamics (see the discussion in \Cref{sec:scaling}), the numbers do not agree within uncertainties.
This indicates that in dimension-1 Betti numbers the dynamics in the infrared is not well-separated from ultraviolet dynamics for averaging over every $8^3$ or $16^3$ blocks.
Additional potential sources of uncertainties have been discussed at the end of \Cref{sec:energytransport}.
 
	\bibliography{main}
	
\end{document}